 \documentclass[preprint]{elsarticle}

\usepackage{graphics}
\usepackage{graphicx}
\usepackage{epsfig}
\usepackage{amssymb}
\usepackage{amsthm}
\usepackage{amsmath}
\usepackage[linesnumbered,ruled,vlined,commentsnumbered]{algorithm2e}
\usepackage{float}
\usepackage[usenames, dvipsnames]{xcolor}
\usepackage{tikz}
\usepackage{subfigure}

\usepackage[misc]{ifsym}
\usepackage{graphics}
\usepackage{epsfig}
\usepackage{amssymb}
\usepackage{amsmath}
\usepackage{stmaryrd}
\usepackage[linesnumbered,ruled,vlined,commentsnumbered]{algorithm2e}
\usepackage{float}
\usepackage{color}
\usepackage{mathabx}
\usepackage{arydshln,leftidx,mathtools}
\usepackage{subfigure}
\usepackage{epsf} 
\usepackage{psfrag}

\def\b0{\mbox{\boldmath $0$}}

\newcommand{\bsigma}{\mbox{\boldmath $\sigma$}}

\newcommand{\bPsi}{\mbox{$\mathbf \Psi$}}

\newcommand{\bepsilon}{\mbox{\boldmath $\varepsilon$}} 

\def\bxi{\mbox{\boldmath$\xi$}}
\newcommand{\bA}{\mbox{\bf A}}

\newcommand{\bC}{\mbox{\bf C}}
\newcommand{\bD}{\mbox{\bf D}}

\newcommand{\bF}{\mbox{\bf F}}
\newcommand{\bG}{\mbox{\bf G}}

\newcommand{\bI}{\mbox{\bf I}}

\newcommand{\bK}{\mbox{\bf K}}

\newcommand{\bM}{\mbox{\bf M}}

\newcommand{\bU}{\mbox{\bf U}}

\newcommand{\bu}{\mbox{\bf u}}
\newcommand{\bx}{\mbox{\bf x}}

\newlength{\figwidth}
\newlength{\figwidthfull}
\newlength{\figwidthhalf}
\newlength{\figwidththird}
\newlength{\figwidthquad}
\newlength{\figspace}

\journal{  }

\begin{document}

\begin{frontmatter}

\title{Data-Informed Decomposition for Localized Uncertainty Quantification of Dynamical Systems}

 \author{Waad Subber\corref{cor1}}
\ead{Waad.Subber@ge.com} 

\cortext[cor1]{Corresponding author}

\author{Sayan Ghosh,  Piyush Pandita, Yiming Zhang, Liping Wang}
\address{Probabilistic Design and Optimization Group,\\
GE  Research,\\
1 Research Circle, Niskayuna, NY 12309, USA}

\begin{abstract}
Industrial dynamical systems often exhibit multi-scale response due to  material heterogeneities, operation conditions and complex environmental loadings. In such  problems, it is the case that the smallest length-scale of the systems dynamics controls the numerical resolution required to effectively resolve the embedded physics.  In practice however, high numerical resolutions is only required in a confined region of the system where fast dynamics or localized material variability are exhibited, whereas a coarser discretization can be sufficient in the rest majority of the system. To this end, a unified computational scheme with uniform spatio-temporal resolutions for uncertainty quantification can be very computationally demanding. Partitioning the complex dynamical system into smaller easier-to-solve problems based of the localized dynamics and material variability can reduce the overall computational cost. However, identifying the region of interest for high-resolution and intensive uncertainty quantification can be a  problem dependent. The region of interest can be specified based on the localization features of the solution, user interest, and correlation length of the random material properties. For problems where a region of interest is not evident, Bayesian inference can provide a feasible solution. In this work, we employ a Bayesian framework to update our prior knowledge on the localized region of interest using measurements and system response. To address the computational cost of the Bayesian inference, we construct a Gaussian process surrogate for the forward model. Once, the localized region of interest is identified,  we use polynomial chaos expansion to propagate the localization uncertainty. We demonstrate our framework through numerical experiments on a three-dimensional elastodynamic problem.
   
\end{abstract}


\end{frontmatter}
 
\section{Introduction} 
  
With the increase in demand for high-performance and highly-efficiency systems, the complexity of industrial design and manufacturing process is increasing proportionally; exposing many opportunities for novel technologies as well as many associated technical challenges. For example, advancements in the design of composite structures allows us to reduce weight, advancements in additive manufacturing enables us to reduce cost, and optimal computational material design pushes the boundary in the discovery of new alloys with desirable electro-mechanical properties. Introducing a new technology typically happens at the lowest level of the systems hierarchy (e.g., at the parts or sub-component levels). Extending the new technologies to the system level requires rigorous testing. For example, in 1980’s composite material was only used for limited components of an aircraft (i.e,. the wing and tail~\cite{dutton2004composite}). Recently however, after multiple test flights, about $50\%$ of the materials used in the Boeing 787 Dreamliner are composite materials~\cite{mrazova2013advanced}. 

In the industrial setting, the process of adaptation of a new technology can be accelerated by a  proper assessment of uncertainty at various aspects of the product 's life cycle spanning the design, manufacturing and maintenance stages. For example, at the design stage of  an aircraft wing rib, it is crucial to consider the effect of uncertainty in the material and operation conditions on the safety factor and aeroelastic dynamics of the wing~\cite{pettit2004uncertainty}. At the manufacturing stage, it is important to consider the impact of manufacturing uncertainties on quality control~\cite{ munk2004determinant} and non-destructive testing~\cite{katunin2015damage}. The maintenance stage requires a holistic assessment of the effect of measurement uncertainty on the static and dynamic responses of the wing during structural health monitoring~\cite{diamanti2010structural}. 

Quantifying uncertainty at the system level often requires a physics-based computational model for the entire structure. However, in structures such as an aircraft wing, traditional computational models may become too complex and costly for simulating  the multi-scale dynamical response especially due to material heterogenity at the sub-component level.
The effect of the sub-component on the entire structure depends on the size, location and loading conditions of the part. It is therefore, necessary to consider a different level of fidelity for the analysis of the sub-components in order to reduce the cost and complexity of uncertainty quantification. To this end, the concept of localized uncertainty propagation for dynamical systems having muti spatio-temporal scales can be utilized to address such issues~\cite{subber2017asynchronous,subber2018uncertainty,subber2016asynchronous}.

In this work, we consider assessing the effect of localized uncertainty in a region of interest within the entire structure. For structures composed of distinct parts that can be clearly identified, the localized region for uncertainty propagation may become obvious. When the distinguished components of a structure are not clear, measurement data can be utilized within Bayesian framework to identify the localized region of interest.
The Bayesian paradigm integrates the \emph{domain-knowledge}, physics-based computational models and observational data in one framework to update the current state of knowledge~\cite{smith2013uncertainty,gelman2014bayesian}. The Bayesian methods offer two major advantages namely: a) allow quantification of epistemic uncertainty under limited-data, and b) retain physical sense for the parameters and the quantity-of-interest. Conditioning apriori physics beliefs on the available data, Bayes rule provides aposteriori distribution on the model parameters. A robust method to estimate the posterior distribution in the  Bayesian inference (i.e. sampling values of the model parameters from the posterior probability) is Markov Chain Monte Carlo (MCMC)~\cite{chen2012monte,gelman2013bayesian}. Estimating the posterior probability density function in the Bayesian method requires solving the forward model many times, which may become challenging for limited computing resources. This issue is often addressed by building a data-driven probabilistic surrogate model using Gaussian process (GP) regression~\cite{williams2006gaussian}. Constructing a GP model requires executing the forward model only few number of times. The GP models are non-parametric and Bayesian in nature, and they provide  uncertainty bound on their predictions. Nevertheless, for problems with stochastic field representation of the variability in the propagation media, uncertainty quantification using GP models may become challenging for general non-Gaussian description of the underlying random variability. Polynomial Chaos (PC), on the other hand, provides an effective framework to represent and propagate  an arbitrary random variable through complex computational models~\cite{ghanem2003stochastic,le2010spectral}. In PC, the response of the physical model is represented as spectral expansion in a polynomial series with basis function being orthogonal with respect to the probability density function of the underlying random variables of the propagation media.

The rest of this work is organized as follows: in Section~\ref{Methodology}, we provide the problem statement and the associated mathematical formulations.  Our numerical demonstrations for the mathematical framework are provided in Section~\ref{NumericalExamples}. We provide the conclusions of the current work in Section~\ref{Conclusion}.

\section{Methodology}
\label{Methodology}
In this  section, we present the mathematical framework of our approach for data-driven partitioning scheme for localized uncertainty quantification. In particular, in Subsection~\ref{Bayesianinference}, we introduce the problem statement in the Bayesian setting. As mentioned previously, for problems where the localized region of interest is not defined explicitly, we rely on measurement data of a response quantity (aided by a computational model) to infer the localized region of interest. The Bayesian framework requires a computational model (the forward problem) to estimate the response of the system for a given set of the input parameters that to be inferred. Consequently, in Subsection~\ref{TheForwardProblem} we discuss the stochastic elastodynamic problem and its finite element discretization. Estimating the localized region of interest in the Bayesian setting necessitates many  solutions to the stochastic elastodynamic problem which can become computationally demanding. Furthermore, it is worth noting that for the Bayesian calculation the entire solution field of the stochastic elastodynamic problem is not required, only a realization at the measurement location is needed. Thus, a surrogate model for the realization of the response can be used to reduce the computational cost of the Bayesian framework. In Subsection~\ref{GP}, we present the Gaussian process surrogate to emulate the solution to the stochastic elastodynamic problem with less cost.  Once the localized region of interest is estimated, a confined uncertainty representation of the material properties within the region of interest can be performed. The localized uncertainty is propagated forward through the model in order to estimate its effect on the variability of the response. For this task, we use the polynomial chose expansion for efficient assessment of uncertainty with less cost. The  polynomial chose expansion is reviewed in Subsection~\ref{PC}.

\subsection{Bayesian Inference}
\label{Bayesianinference}

In Bayesian inference, the prior knowledge is updated to posterior using noisy measurements and the response of a physical model~\cite{smith2013uncertainty,gelman2014bayesian}. The update is based on the Bayes' rule defined as
\begin{equation}
\label{Bayes}
p(\theta|{ \bf d}) = \frac{p(\theta) p({ \bf d}|{\theta})}{p({ \bf d})},
\end{equation}
where $\theta$  is the uncertain parameters to be estimated, $\bf d$ is the measurement of an observable quantity,  $p(\theta|{\bf d})$ is the posterior probability density function, $p(\theta)$ is the prior probability density function, and $ p({ \bf d}|{\theta})$ denotes the likelihood of the observations given the parameter. We assume that the measured data $\bf d$ is generated from a statistical model composed of a physical model ${\mathbb M}(\theta)$ plus an additive measurements noise $\epsilon$ as
\begin{align}
{\bf d} = {\mathbb M}(\theta) +\epsilon.
\end{align} 
\noindent
Here we represent the measurement noise as a Gaussian random variable with unknown variance $\epsilon\sim{\mathcal N}(0,\sigma_n^2)$. For a Gaussian noise, the  likelihood function  becomes
\begin{align}
\label{loglik}
p({ \bf d}|{\theta}) = \frac{1}{\left( 2\pi \sigma_n^2\right)^{-N/2}}\exp\left( - \frac{1}{2\sigma_n^2} \displaystyle\sum_i^N [d_i -  {\mathbb M}(\theta_i)]^2\right).
\end{align}
\noindent
The task in hand is to utilize the measurement $\bf d$  and the physical model  ${\mathbb M}(\theta)$ to estimate the system parameters $\theta$   that best satisfy Eq.\eqref{Bayes}. The process requires many executions to the physical model ${\mathbb M}(\theta)$, which can be computationally expensive. It is often, the expensive computational model is emulated by a simpler {\it easy-to-evaluate} model that can estimate the response with a quantified accuracy as:
\begin{align}
{\mathbb M}(\theta) ~\simeq {\mathcal M}(\theta),
\end{align}
\noindent
where ${\mathcal M}(\theta)$ denotes the surrogate model that is constructed using a limited runs of the physical model ${\mathbb M}(\theta)$. In our work, we represent ${\mathcal M}(\theta)$ as  Gaussian process surrogate model~\cite{rasmussen2003gaussian}.
\noindent
Once we constructed and validate the surrogate model, the parameterization of the localized features $\theta$  is estimated using Bayes' rule evaluated by Markov Chain Monte Carlo (MCMC) sampling technique. Having the localized region of interest identified, a localized uncertainty quantification of the confined variability of the material properties can be performed efficiently using polynomial chaos expansion~\cite{ghanem2003stochastic}.

\subsection{The Forward Problem}
\label{TheForwardProblem}
\begin{figure} [H]
\centering
\subfigure[Spatial domain]{\includegraphics[width=0.85\figwidthhalf,keepaspectratio]{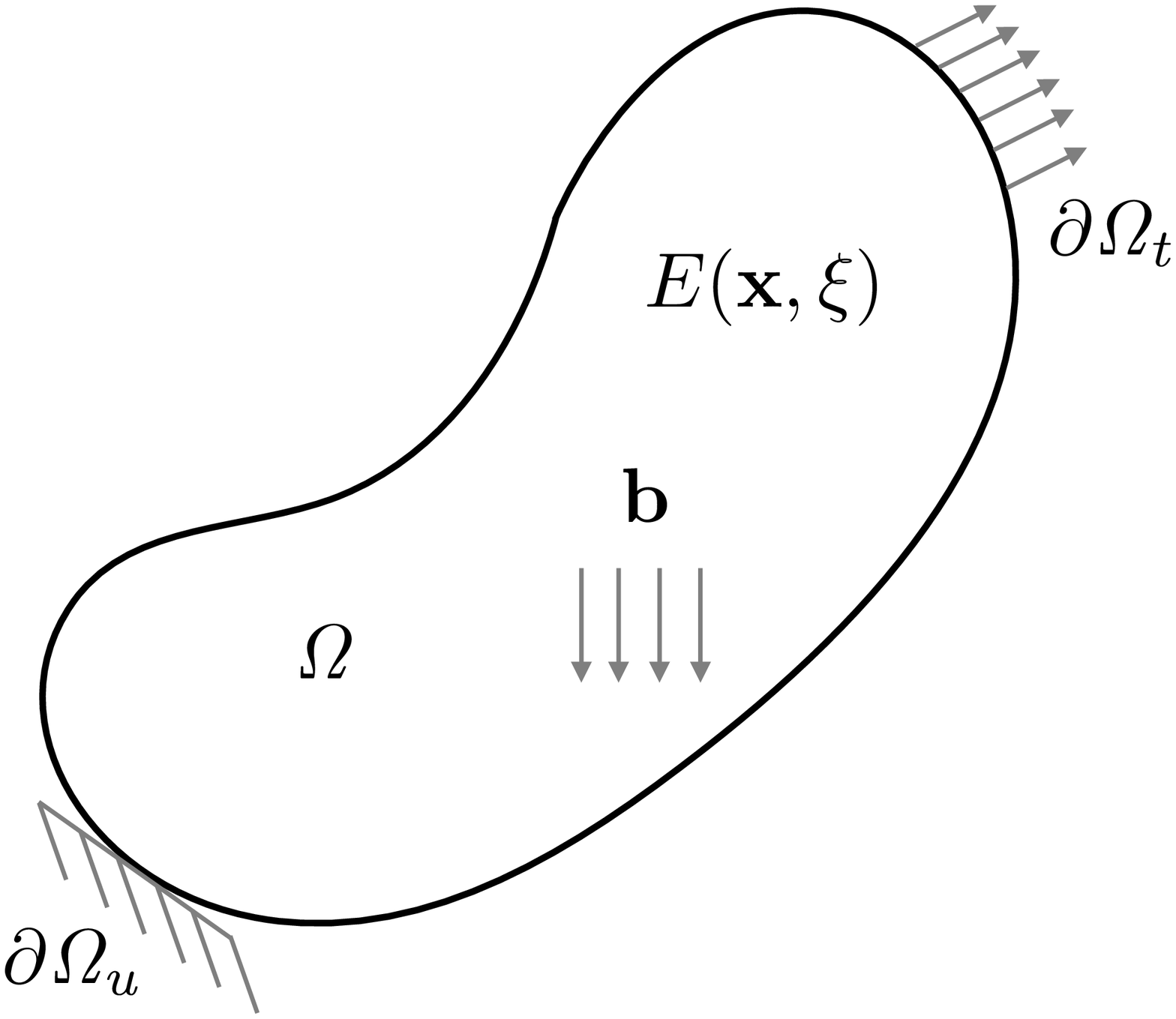}}
\subfigure[Domain decomposition.]{\includegraphics[width=0.75\figwidthhalf,keepaspectratio]{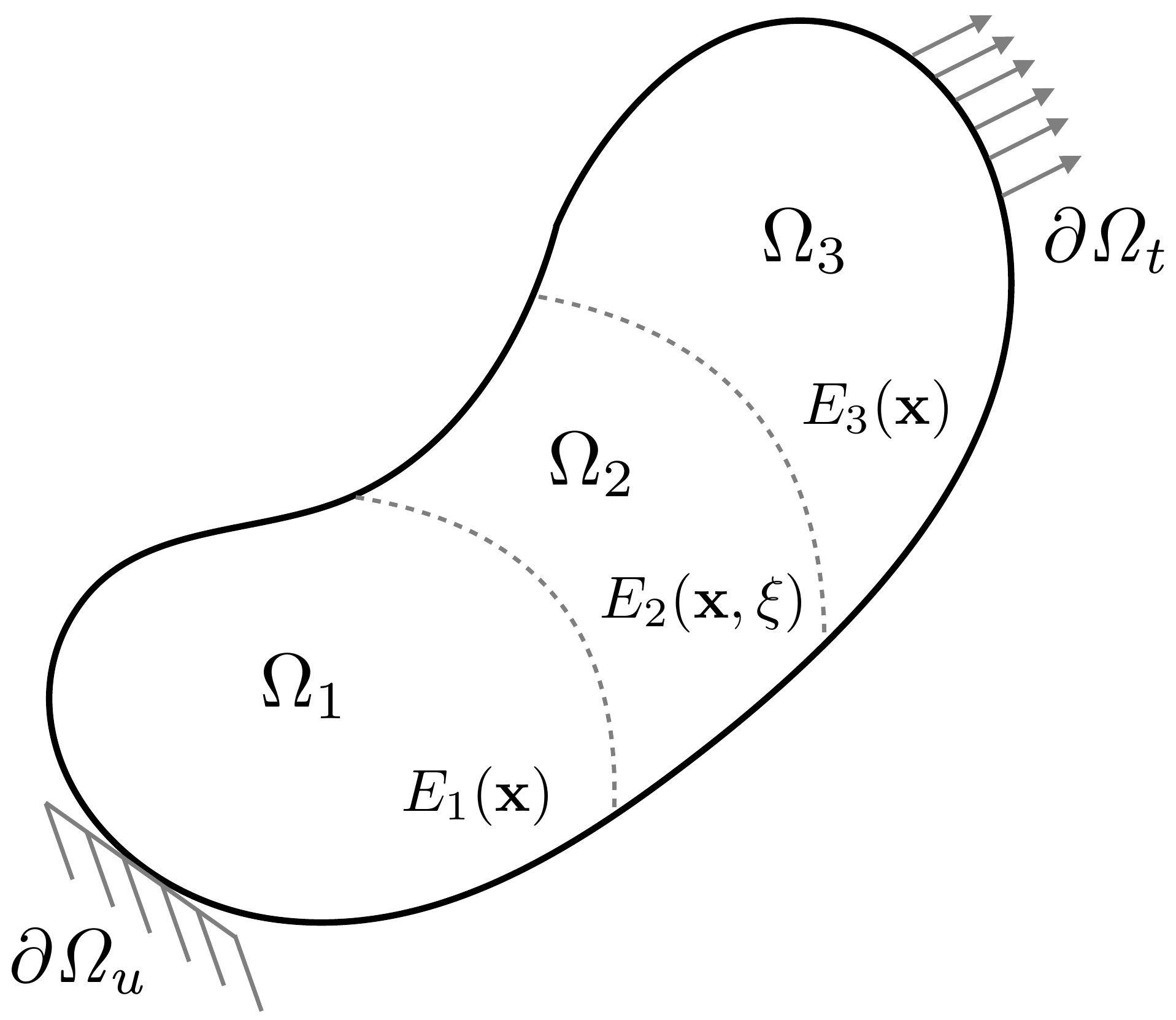}}
 \caption{ An arbitrary computational domain $\Omega$ with a random material property (i.e., $E({\bf x},\bxi)$) and its partitioning into  non-overlapping subdomains. The partitioning is based on material variability.} 
 \label{domain}
 \end{figure}

In this section, we give a brief summary of the mathematical formulation to the linear stochastic dynamical system considered in this work. The framework is presented for localized uncertainty propagation work-flow, whereby the decomposition of the physical domain is based on the variability of the material properties. Consequently, we consider an arbitrary physical domain  $\Omega \in \mathbb{R}^d$ with $\partial \Omega$ being its boundary as shown in Fig.~(\ref{domain}-a), and define the following problem:

Find a random function ${\bf u}({\bf x},t,\bxi):\Omega \times [0,T_f]\times \Xi \rightarrow  \mathbb{R}$, such that the following equations hold
\begin{align}
\label{EqOfMotion}
\begin{array}{rlcllll}
\rho(\xi) \ddot{\bf u}(\bx,t,\bxi)=&\nabla \cdot \bsigma +{\bf b} &\text{in} &\Omega&\times&[0,T_f]~\times \Xi,\\
{\bf u}(\bx,t,\bxi) =&\bar{{\bf u}}    &\text{on}& \partial \Omega_u&\times&[0,T_f]~\times \Xi,\\
\bsigma \cdot {\bf n} =&\bar{{\bf t}} &  \text{on}& \partial \Omega_t&\times&[0,T_f]~\times \Xi,\\
{\bf u}(\bx,0,\bxi) =&{{\bf u}}_0   & \text{in}& \Omega&\times& \Xi,\\
\dot{{\bf u}}(\bx,0,\bxi) =&\dot{{\bf u}}_0   & \text{in}& \Omega&\times& \Xi,
\end{array}
\end{align}
\noindent
where  $\rho(\xi)$ is the mass density, $\bsigma$ is the stress tensor, ${\bf u}$ is the displacement field, $\bf b$ is the body force per unit volume, $\bar{{\bf u}}$ is the prescribed displacement on $\partial \Omega_u$, $\bar{{\bf t}}$ is the prescribed traction on $\partial \Omega_t$,
$\bf n$ is a unit normal to the surface, and ${{\bf u}}_0$ and $\dot{{\bf u}}_0$ are the initial displacement and velocity, respectively. Here,  we define the stochastic space by ($\Theta,\Sigma,P)$, where $\Theta$ denoting the sample space, $\Sigma$ being the $\sigma$-algebra of  $\Theta$, and $P$ representing an appropriate probability measure. The stochastic space is paramatrized by a finite set of standardized identically distributed random variables $\bxi = \{ \xi_i(\theta)\}_{i=1}^M$, where $\theta \in \Theta$. The support of the random variables is defined as $\Xi = \Xi_1 \times \Xi_2 \times \cdots \Xi_M \in \mathbb{R}^M$ with a  joint probability density function given as $p(\bxi) =p_1(\xi_1)\cdot p_2(\xi_2)\cdots p_M(\xi_M)$. 

For linear isotropic elastic martial, the constitutive relation between the stress  and strain tensors is given by:
\begin{equation}
\bsigma = \lambda(\bxi)\rm{tr}(\bepsilon) \bI + 2\mu(\bxi) \bepsilon,
\end{equation}
\noindent 
where $\lambda(\bxi)$ and $\mu(\bxi)$ are the Lema\'e's parameters, $\bI$ is an identity tensor and $\bepsilon$ is the symmetric strain tensor defined as
\begin{equation}
\bepsilon =\frac{1}{2} \left( \nabla \bu + \nabla \bu^T \right).
\end{equation}
\noindent
For a random  Young's modulus $E(\bx, \bxi) $ and deterministic Poisson's ratio $\nu$ ,  the Lema\'e's  parameters can be expressed as 
\begin{align}
\lambda(\bxi) &= \frac{E(\bx, \bxi) \nu}{(1+\nu)(1-2\nu)},\quad
\mu(\bxi) = \frac{E(\bx, \bxi)}{2(1+\nu)}.
\end{align}

We consider the case that uncertainty steams from a localized variability in a confined region within the physical domain. For example as shown in Fig.~(\ref{domain}-b), the variability in the quantity of interest can be attributed to the material random properties within the subdomain $\Omega_2$. The artificial martial boundaries shown in Fig.(~\ref{domain}-b) for subdomain $\Omega_2$ is estimated using Bayesian inference. Localizing random variability in the neighborhood of the quantity of interest reduces the computational cost of uncertainty propagation in problems where a region of interest can be specified. Depending on the interest in the region,  each subdomain can have its local uncertainty representation and the corresponding mesh and time resolutions. As a results the Asynchronous Space-Time Domain Decomposition Method with Localized Uncertainty Quantification (PASTA-DDM-UQ)~\cite{subber2017asynchronous,subber2018uncertainty,subber2016asynchronous} can be utilized. In PASTA-DDM-UQ, spatial, temporal and material decompositions are considered. In this work however, we only consider material decomposition and apply non-intrusive approach for uncertainty propagation.

Consequently, let the physical domain  $\Omega$ be partitioned based on the martial variability into $n_s$ non-overlapping subdomains $\Omega_s, 1\le s\le n_s$ as shown in Fig.~(\ref{domain}-b) and such that:
\begin{align}
\label{Decomp}
\Omega &= \displaystyle\bigcup_{s=1}^{n_s} \Omega_s, \quad \Omega_s \displaystyle\bigcap  \Omega_r=\emptyset~\text{for} ~s\neq r,\quad 
 \Gamma= \displaystyle\bigcup_{s=1}^{n_s} \Gamma_s, \quad \Gamma_s = \partial\Omega_s \backslash \partial\Omega.
\end{align}

According to the decomposition in Eq.\eqref{Decomp}, the stochastic dynamical problem in Eq.\eqref{EqOfMotion} can be transformed into the following  minimization problem: 

Find a random function ${\bf u}({\bf x},t,\bxi):\Omega \times [0,T_f]\times \Xi \rightarrow  \mathbb{R}$, such that
\begin{align}
\label{TotalLagrangian}
& {\mathcal L} ( \bu, \dot{\bu}) = \displaystyle\sum_{s=1}^{n_s} \left({\mathcal T}_s(\dot{\bu}) -{\mathcal V}_s(\bu)\right) \rightarrow \text{\rm min}, \quad s=1,\cdots,n_s,
\end{align}
\noindent
where ${\mathcal L} ( \bu, \dot{\bu})$ is the Lagrangian of the system,  ${\mathcal T}_s(\dot{\bu})$ denotes the  subdomain kinetic energy and ${\mathcal V}_s(\bu)$ is the subdomain potential energy defined as:

\begin{align}
\label{Minimization}
{\mathcal T}_s(\dot{\bu}) &=  \int_{\Xi}    \int_{\Omega_s}\frac{1}{2} \rho_s(\xi) {\dot \bu} \cdot {\dot \bu} \, {\rm d} \Omega  {\rm d}\Xi, \\
{\mathcal V}_s(\bu) &= \int_{\Xi} \left(\int_{\Omega_s}  \frac{1}{2} \bepsilon : \bsigma_s \, {\rm d}\Omega 
+\int_{\Omega_s} \bu \cdot {\bf b}_s   \, {\rm d} \Omega+  \int_{\partial\Omega_t} \bu \cdot \bar{\bf t}_s   \, {\rm d} \Gamma   \right)  {\rm d}\Xi,
\end{align}

The Hamilton's principle with a dissipation term reads
\begin{align}
\label{Hamilton}
\int_0^{T_f} \left( \delta {\mathcal L} - \frac{\partial {\mathcal Q}}{\partial \dot{\bepsilon}  } : \delta {\bepsilon} \right)  {\rm d t}=0,
\end{align}
\noindent
where  $\delta {\mathcal L}$ is the first variation of the augmented Lagrangian defined as
\begin{align}
\label{Variation}
\nonumber
&\delta {\mathcal L} =\displaystyle\sum_{s=1}^{n_s}  \int_{\Xi}  \left( \int_{\Omega_s} \rho_s(\xi) \delta {\dot \bu} \cdot {\dot \bu} \, {\rm d} \Omega - \int_{\Omega_s} \delta \bepsilon : \bD_s(\bxi): \bepsilon \, {\rm d}\Omega +\right. \\
& \left. \int_{\Omega_s} \delta \bu \cdot {\bf b}_s   \, {\rm d} \Omega+   \int_{\partial\Omega_t} \delta \bu \cdot \bar{\bf t}_s   \, {\rm d} \Gamma  \right) {\rm d}\Xi,
\end{align}
\noindent 
here we define $\bD_s(\bxi)$ as the uncertain linear elasticity tensor. The dissipation function ${\mathcal Q} ( \dot{\bu})$ in the Hamilton is defined as
\begin{align}
\label{dissipation}
&{\mathcal Q} ( \dot{\bu}) = \sum_{s=1}^{n_s}  \frac{1}{2} \int_{\Xi}  \int_{\Omega_s}  \dot{\bepsilon} :{\widehat{\bD}}_s: \dot{\bepsilon} \, {\rm d}\Omega {\rm d}\Xi, \quad s=1,\cdots,n_s,
\end{align}
\noindent
where ${\widehat{\bD}}_s$ is the damping tensor assumed to be deterministic.
\noindent
Substituting $\rm{Eqs}.~(\ref{dissipation}~and~\ref{Variation})$ into the  Hamilton's principle   Eq.~\eqref{Hamilton} gives the following  stochastic equation of motion for  a typical subdomain $\Omega_s$
\begin{align}
\label{ContinuumEqMotion}
&\int_{\Xi} \int_{\Omega_s} \rho_s(\xi) \ddot{\bu} \cdot \delta\bu \, {\rm d} \Omega \,{\rm d}\Xi
+\int_{\Xi} \int_{\Omega_s} \dot{\bepsilon}:\widehat{\bD}_s : \delta\bepsilon \, {\rm d} \Omega \, {\rm d}\Xi+\int_{\Xi} \int_{\Omega_s} \bepsilon :\bD_s(\bxi): \delta\bepsilon \, {\rm d} \Omega \, {\rm d}\Xi\\
\nonumber 
&=\int_{\Xi}  \int_{\Omega_s} \delta\bu \cdot {\bf b}_s \,{\rm d} \Omega\, {\rm d}\Xi
+ \int_{\Xi}  \int_{\partial\Omega_t} \delta\bu \cdot \bar{\bf t}_s   \, {\rm d} \Gamma\, {\rm d}\Xi.
\end{align}

In the next section, we describe the finite element discretization of the weak form Eq.\eqref{ContinuumEqMotion}.

\subsubsection{Spatial and Temporal  Discretizations}
\label{FEM}
Let the spatial domain $\Omega$ be triangulated with finite elements of size $h$ and let the associated finite element subspace be defined as $\mathcal X_h\subset H_0^1(\Omega)$, then an approximate  deterministic finite element  solution can be expressed as
\begin{align}
\label{FEMd}
{\bf u}^h=\sum_i^{n_i} {\bf N}_i(\bx) \tilde{\bf u}^i(t).
\end{align}
\noindent
Substituting the discrete field, Eq.\eqref{FEMd} in the weak form Eq.\eqref{ContinuumEqMotion} gives the following semi-discretized stochastic equation of motion :

\begin{align}
\label{discreteEqMotion2}
\int_{\Xi}  \left( {\bf M} \ddot{\bf u}(t) +{\bf C} \dot{\bf u}(t)   +{\bf K} {\bf u}(t)  \right) {\rm d}\Xi =\int_{\Xi}  {\bf F}(t) {\rm d}\Xi.
\end{align}
\noindent
We drop the nodal finite element marks (tilde) for brevity of the representation and define the following matrices:
\begin{align*}
{\bf M} &=  \displaystyle\sum_{s=1}^{n_s}  \int_{\Omega_s} \rho_s{\bf N}^T {\bf N} {\rm d} \Omega , \quad
{\bf C} = \displaystyle\sum_{s=1}^{n_s}  \int_{\Omega_s} {\bf B}^T \widehat{\bD}_s{\bf B} {\rm d} \Omega , \\
{\bf K} &=  \displaystyle\sum_{s=1}^{n_s} \int_{\Omega_s} {\bf B}^T {\bf D}^i_s{\bf B} {\rm d} \Omega , \quad  {\bf F}(t) = \displaystyle\sum_{s=1}^{n_s} \left(\int_{\Omega_s} {\bf b}_s^T {\bf N}{\rm d} \Omega + \int_{\partial\Omega_s} {\bf \bar t}_s^T {\bf N} {\rm d}\Gamma \right) .
\end{align*}
\noindent
Here, ${\bf B}$ is the displacement-strain matrix.
\noindent
For time discretization, we use the Newmark  time integration scheme to advance the stochastic system one time step as
\begin{align}
\label{Newmark11}
\dot{\bf u}^{k+1}&= \dot{\bf u}^{k} +(1-\gamma) \Delta t \ddot{\bf u}^k+\gamma\Delta t \ddot{\bf u}^{k+1},\\
\label{Newmark21}
{\bf u}^{k+1} &= {\bf u}^{k}+ \Delta t \dot{{\bf u}}^k  +\left(\frac{1}{2}-\beta\right) \Delta t^2 \ddot{{\bf u}}^k+\beta \Delta t^2 \ddot{{\bf u}}^{k+1},
\end{align}
where $\gamma$ and $\beta$ are the integration parameters, and $\Delta t=\frac{T_{f}-T_0}{n_{t}}$. Substituting he Newmark scheme into the semi-discretized stochastic equation of motion Eq.\eqref{discreteEqMotion2}, gives the following fully discretized 
 linear system for a give realization of the random vector $\bxi $:

\begin{align}
\label{DiscretSaddlePointMatrixForm}
{\bA} (\bxi){\bU}^{k+1}(\bxi)=
{\bF}^{k+1}-{\bG}{\bU}^k(\bxi)
\end{align}
\noindent
where for compact representation, we define
\begin{align*}
{\bA} (\bxi)&=\left[
\begin{array}{ccc}
\bM(\xi) & \hspace{2mm} \bC & \hspace{2mm} \bK (\bxi) \\
-\gamma \Delta T \bI & \hspace{2mm} \bI & \hspace{2mm} \b0\\
-\beta \Delta T^2\bI & \hspace{2mm} \b0 & \hspace{2mm} \bI
\end{array}
\right],
\quad
{\bG}=\left[
\begin{array}{ccc}
\b0 & \hspace{2mm} \b0 & \hspace{2mm} \b0  \\
-(1-\gamma) \Delta T\bI  & \hspace{2mm} -\bI &  \hspace{2mm} \b0\\
-(\frac{1}{2}-\beta)\Delta T^2\bI  &  \hspace{2mm} -\Delta T\bI &  \hspace{2mm} -\bI
\end{array}
\right],\\
{\bU}(\bxi)&=\left\{
\begin{array}{c}
{\ddot \bu}(\bxi)\\
{\dot \bu}(\bxi)\\
\bu(\bxi)
\end{array}
\right \},\quad
{\bF}=\left \{
\begin{array}{c}
{\bf f}\\
\b0\\
\b0
\end{array}
\right \}.
\end{align*}
\noindent
For the data-driven decomposition approach, many solutions to the forward problem Eq.~\eqref{DiscretSaddlePointMatrixForm} are required in estimating the appropriate decomposition for localized uncertainty propagation. 
To mitigate the computational cost involved with identifying the underlying localized region of interest, a Gaussian Process (GP)  surrogate model is utilized as explained in the next section.

\subsection{Surrogate  Modeling} 
\label{GP}

The Gaussian Process (GP) surrogate model is widely used for engineering problems as a cost-effective alternative to costly computer simulator~\cite{ghosh2020efficient}. 
In the authors' previous work~\cite{ghosh2020advances}, a fully-Bayesian industrial-level implementation for GP-based metamodeling and model calibration has been exhaustively covered.
This implementation, called GE's Bayesian hybird modeling (GEBHM), has been rigorously tested and validated on numerous benchmark problems and the impact of using Bayesian surrogate modeling has been demonstrated on several challenging industrial problems.
In GP for dynamical systems, we consider $ \mathcal{ D}= \{ ({\bf x}_i,{\bf y}_i)~|~i =1,2,\cdots,N\}$ to be a set of training data consists of $N$ samples, where ${\bf x}_i \in \mathbb{R}^d$ represents the input sample $i$, and  ${\bf y}_i$ is the corresponding output vector of size $n_T$. For time-series data, the output is observed at a sequence of time steps $t_j \in [t_1,t_2,\cdots,t_{n_T}]$. We concatenate all the input and output into the design matrix $\mathcal{X}$ and the corresponding observation matrix $\mathcal Y$, respectively as:

\begin{align}
\label{training1}
\mathcal{X}= \left[
\begin{array}{cc}
t_1 & {\bf x}_1\\
\vdots & \vdots\\
t_{n_T} & {\bf x}_1\\
\vdots & \vdots\\
t_1 & {\bf x}_N\\
\vdots & \vdots\\
t_{n_T} & {\bf x}_N\\
\end{array}
\right],~\quad\quad
\mathcal{Y} = \left[
\begin{array}{c}
{y}_1^1\\
\vdots \\
y_{n_T}^1\\
\vdots \\
{y}_1^N\\
\vdots \\
{y}_{n_T}^N\\
\end{array}
\right],
\end{align}
\noindent
where $y_j^i$ is the response at time $t_j$ for the input parameters ${\bf x}_i$. The sizes of the design matrix ${\mathcal X}$ and the observation matrix ${\mathcal Y}$ are $(N\times n_T) \times (d+1)$ and $(N\times n_T) \times 1$, respectively. In compact form, the training dataset  ($\mathcal X, Y$) can be rewritten as:
\begin{align}
\mathcal{X} = \left[ 
\begin{matrix}
{\bf 1}_N \otimes {\bf T} & {\bf X} \otimes {\bf 1}_{n_T}
\end{matrix}
\right],\quad\quad
\mathcal{Y} = {\rm vec} ({\bf Y}),
\end{align}
\noindent
where ${\bf 1}_N$ is an identity vector of size $N$, ${\bf  X}=[{\bf x}_1,\cdots, {\bf x}_N]^T$, ${\bf T}=[t_1,\cdots t_{n_T}]^T$,  ${\bf 1}_{n_T}$ is an identity vector of size $n_T$,  ${\bf Y}=\left[ \begin{matrix} {\bf y}^1 & \cdots& {\bf y}^N \end{matrix} \right]$ and ${\bf y}^i=[y_1^i,\cdots y_{n_T}^i]^T$. Here the symbols $\otimes$  and $\rm{vec}(\bullet)$ represent Kronecker product and vectorization operators, respectively. Consequently, a general regression model for time-dependent data can be expressed as a function $f({\mathcal X})$ that maps the input ${\mathcal X}$ to time-series observation ${\mathcal Y}$. 
In GP regression, the goal is to infer the  function $f({\mathcal X})$ from noisy observation of the the output ${\mathcal Y}$. To this end, the function $f({\mathcal X})$ is viewed as a random realization of Gaussian processes $ f({\mathcal X})\sim {\mathcal GP} (\mu({\mathcal X}), {\bf K}({\mathcal X},{\mathcal X'}))$, where $\mu({\mathcal X})$ and $ {\bf K}({\mathcal X},{\mathcal X'})$ are the mean and covariance matrix of the process, respectively. Training the GP model can be performed by finding the optimal values to the covariance parameters. Systematically, this is done by maximizing the evidence or the marginal likelihood with respect to the hyperparameter parameters of the kernel. The prediction of the GP for a new input $\bf{ x}_*$, is a Gaussian process with the following posterior mean and covariance
\begin{align}
\label{mp}
\mu({\bf x}_*) &= {\bf k}({\bf x}_*, {\mathcal X}) [{\bf K}({\mathcal X},{\mathcal X'}) + \sigma_n^2 {\bf I}]^{-1}  {\mathcal Y},\\
\label{sp}
\sigma^2({\bf x}_*) &=  {\bf k}({\bf x}_*, {\bf x}_*) -  {\bf k}({\bf x}_*, {\mathcal X}) [{\bf K}({\mathcal X},{\mathcal X'}) + \sigma_n^2 {\bf I}]^{-1} {\bf k}({\mathcal X},{\bf x}_*)
\end{align}

The covariance function in the GP framework encodes the smoothness and it measures the similarity of the process between two points. The covariance function also encodes the prior belief over the regression function to model the measurements. The prior belief can be on the level of the function smoothness, or behavior and trend such as periodicity, for example. Selecting the right covariance kernel can be challenging for time-dependent data and may require a composition of several covariance functions together to model the right behavior of the data. On the other hand, for problems where the training data is given in the form as in Eq.\eqref{training1}, the size of the data may grow exponentially demanding large computational budged. In this case a scaleable framework for the GP regression of large dataset can be exploited to efficiently address the computational cost~\cite{zhang2020remarks}. 

In this work, the ultimate goal of the GP model is to serve as a surrogate to the costly simulation code in the Bayesian inference. Thus, we follow a simplified approach to reduce the cost of building the surrogate~\cite{shabouei2019chemo}. For the case when the time index of  measurement is set a priori and prediction at intermediate time instant is not required, the inter correlation between the time steps can be relaxed. Specifically, the prediction of the model in this case  is always set at the location of the measured data, and the model only considers the correlation among the input variables ${\bf x}_i$. Thus the GP can be constructed on on the subset of the data  (${\bf X}, {\bf Y}$) instead of ($\mathcal X, Y$) as ${\rm GP} (\mu({\bf X}), {\bf K}({\bf X},{\bf X'}))$, where

\begin{align}
\label{training}
{\bf X}= \left[
\begin{array}{c}
 {\bf x}_1\\
\vdots \\
{\bf x}_N
\end{array}
\right],~\quad\quad
\bf{Y} = \left[
\begin{array}{ccc}
{y}_1^1,& \hdots,& {y}_{n_T}^1\\
& \vdots & \\
{y}_N^1,& \hdots,& {y}_{n_T}^N
\end{array}
\right]
\end{align}

\subsection{Polynomial Chaos} 
\label{PC}

The Polynomial Chaos (PC) expansion is based on the spectral decomposition of a stochastic process into deterministic coefficients scaling random functions. In particular, the PC approximates a stochastic process as a linear combination of stochastic orthogonal basis functions as
 
\begin{align}
{\bu}(t,\bxi) &= \displaystyle\sum_{j=0}^{N}  \bPsi_j(\bxi){\bu}_j(t),
\end{align}
\noindent
where $\bPsi_j(\bxi)$ are a set of multivariate orthogonal random polynomials and ${\bu}_j(t),$ are the deterministic projection coefficients. The PC coefficients can be estimated non-intrusively as 

\begin{align}
{\bu}_j(t)&= \frac{ \int_{\Xi}  {\bu}(t,\bxi)  \bPsi_j(\bxi) {\rm d}\Xi}{\int_{\Xi}    \bPsi_j^2(\bxi) {\rm d}{\Xi}  },
\end{align}
\noindent
where $\int_{\Xi} (  \bullet )~  {\rm d}{\Xi} $ denotes the expectation operator with respect to the probability density function of the underlying random variables. The expectation integral can be estimated using random sampling or  deterministic quadrature rule~\cite{feinberg2015chaospy}

\section{Numerical Example}
\label{NumericalExamples}
For the numerical demonstration, we consider the problem of detecting the desired geometry (e.g. localized features) for a given specimen from noisy measurements of its  dynamical response. We paramatrize the geometry by the dimensions of the inner section (the inner length $l_i$ and radius $r_i$) as shown in  Fig.~(\ref{beam}). The inner dimensions  are inferred from noisy measurement of the beam deflection at the mid-span. Once the dimensions are estimated, we perform a localized uncertainty propagation of the material parameters of the inner core.  

\subsection{The Forward Problem}

\begin{figure} [H]
\centering
\includegraphics[width=0.75\figwidthfull,keepaspectratio]{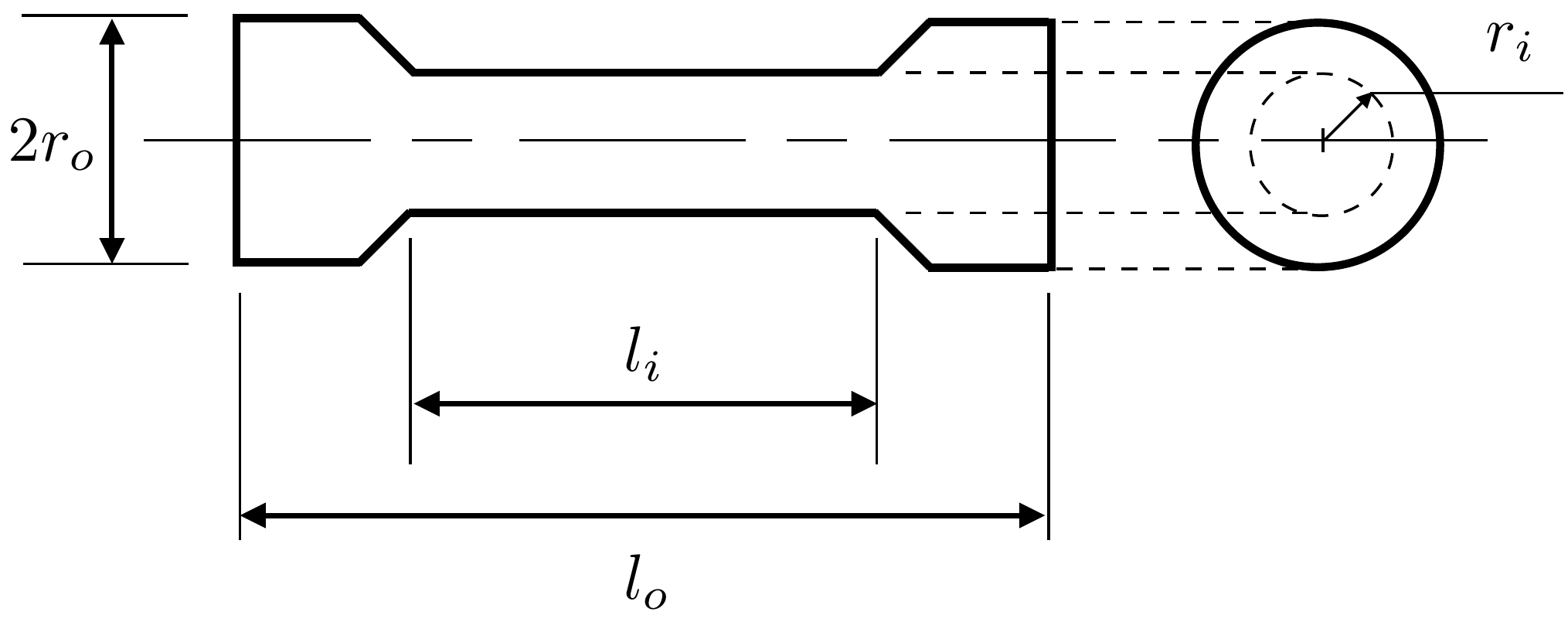}
\caption{Schematic showing a 2D projection of  a typical beam. For the reference case the inner and outer dimensions are  ($l_i = 0.45~{\rm m}, r_i=0.025~{\rm m}$) and  ($r_o = 0.05~{\rm m}~\text{and}, l_o=1.0~{\rm m}$), respectively.}
\label{beam}
 \end{figure}

We  consider a 3-D Aluminum beam with mean elastic properties of $E=70$ GPa, $\nu=0.3$ and $\rho=26.25$ kN/$\rm m^3$.  For the damping representation, we consider Rayleigh damping whereby the damping is assumed as ${\bf C}=\eta_m {\bf M}+\eta_k {\bf K}$. In the numerical implementation, we consider the proportion constants $\eta_m=0$ and $\eta_k=0.001$, and we use stiffness ${\bf K}$ based on the mean properties. We utilize FEniCS for the forward finite element simulations.~\cite{logg2012automated}. Fig.~(\ref{beam}) shows a 2D projection of the beam geometry, whereby we parameterize the inner cylinder by (length $l_i$ and radius $r_i$), and the outer cylinder by (length $l_o$ and radius $r_o$). For the reference case the inner and outer dimensions are  ($l_i = 0.45~{\rm m}, r_i=0.025~{\rm m}$) and  ($r_o = 0.05~{\rm m}~\text{and}, l_o=1.0~{\rm m}$), respectively. The beam is subjected to an impact force defined as:
\begin{align}
F(t,{\bx})  = \left[ 0, 0, {F_0 t /}{t_c} \delta(t-tc) \right ]^T,
\end{align}
\noindent
where $F_0= -5.0$ GN and the ramp time  $t_c = 0.5$ ms. The beam is fixed at both ends and  subjected to zero initial displacement and velocity. The dynamics is integrated  up to $0.01$ s. 

\begin{figure} [H]
\centering
 \subfigure[displacement]{\includegraphics[width=\figwidthhalf,keepaspectratio]{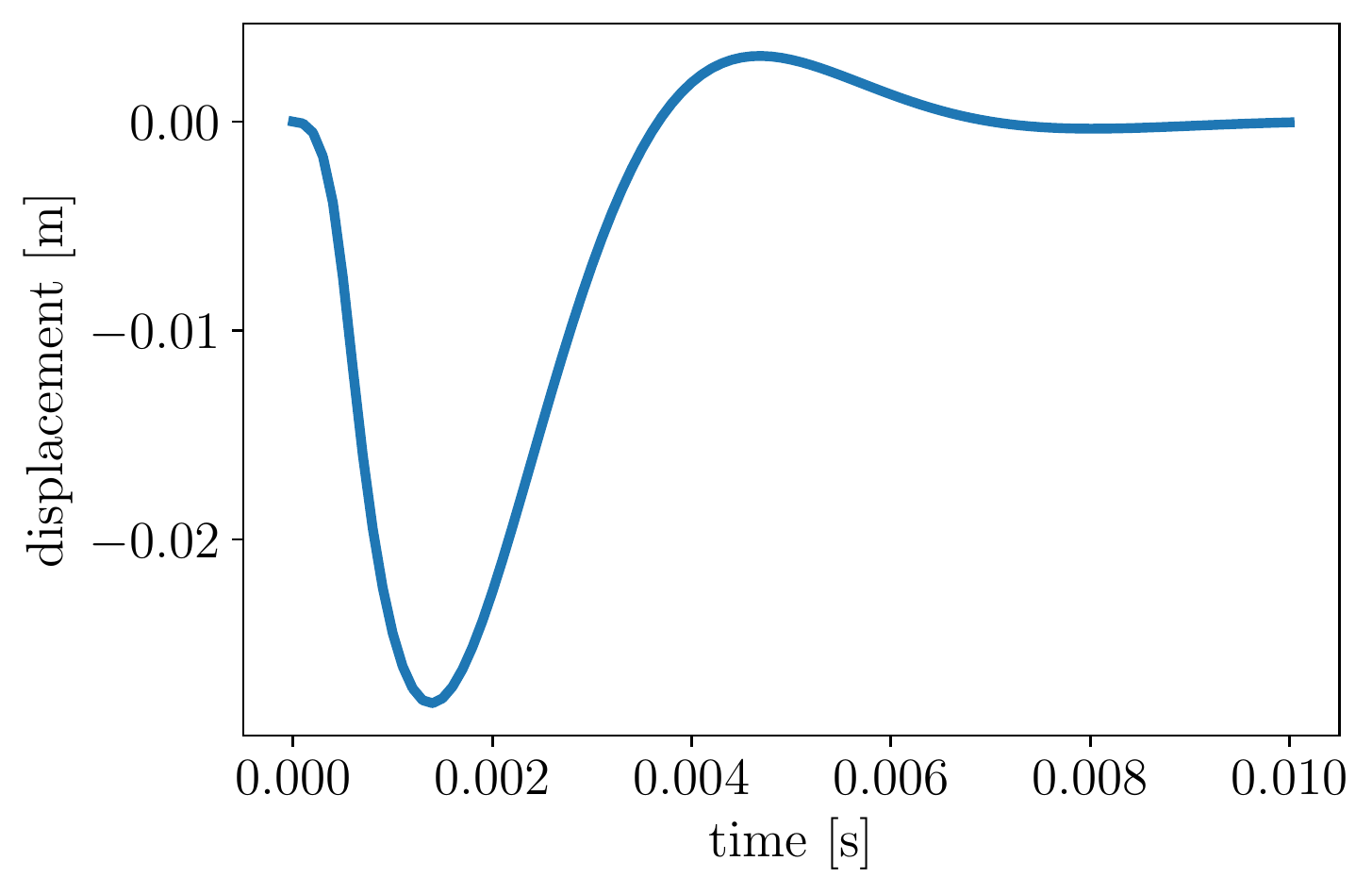}}
  \subfigure[velocity]{\includegraphics[width=\figwidthhalf,keepaspectratio]{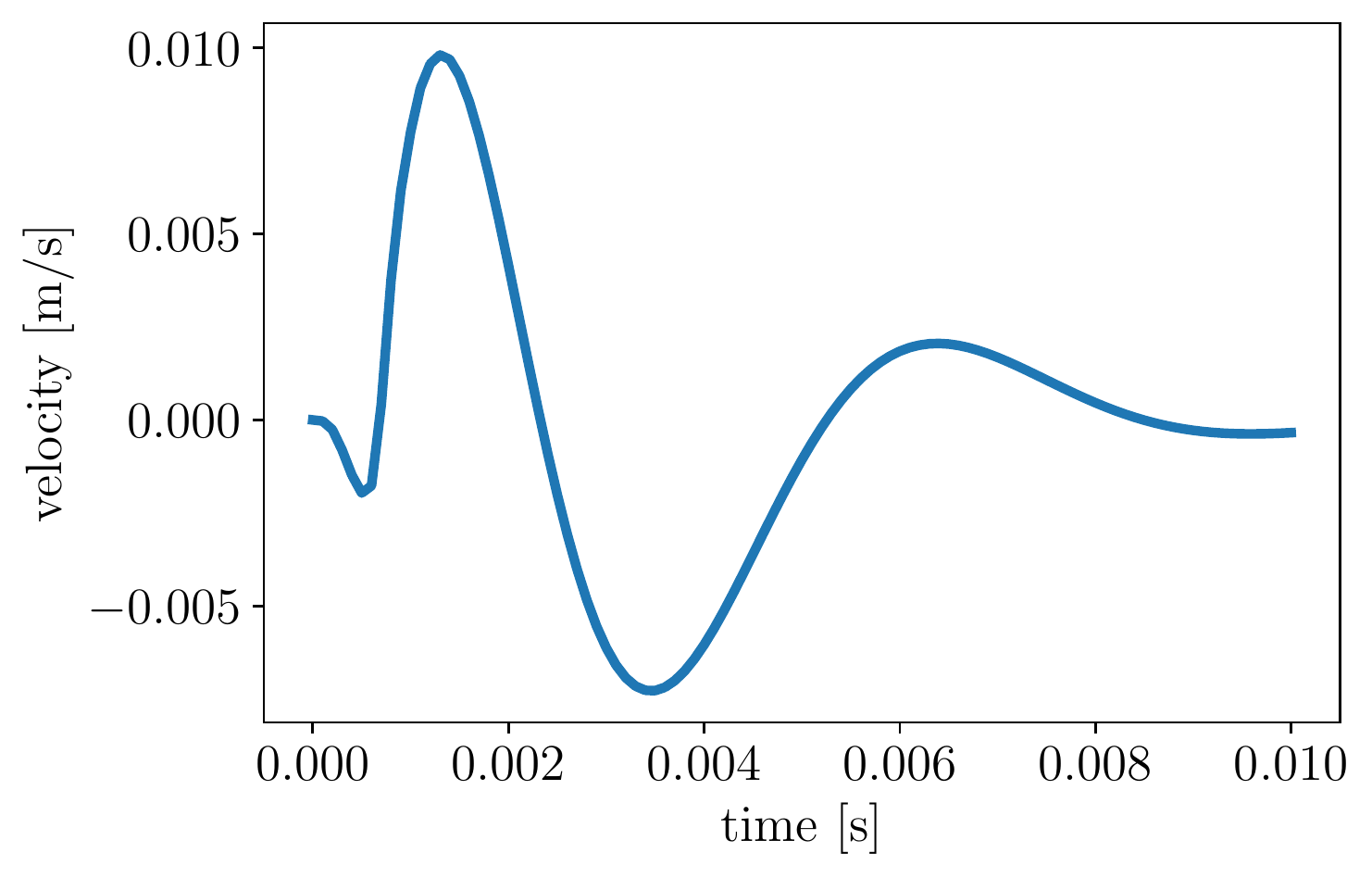}}
\caption{The displacement and velocity at the mid-span of the reference case  $P(0.5,0.0,0.0) [\rm m]$ and using the mean material properties $E=70$ GPa, $\nu=0.3$ and $\rho=26.25$ kN/$\rm m^3$}
\label{case0u0}
 \end{figure}

We consider the vertical deflection at the mid-span  to be the quantity of interest (QoI) in identifying the underlying beam geometry. Fig.~(\ref{case0u0}) shows  he mid-span displacement and velocity for a the reference case.

\subsection{The Surrogate Model}
In order to infer the beam geometry from the QoI, many runs of the forward model (the 3D finite element dynamical code are required. This may cause a computational burden when the computational budget is limited. Thus,  a surrogate model can overcome this issue by utilizing a limited number of a prespecified runs. The design of computer experiments concept can be used to optimally select the required runs~\cite{kristensen2019industrial,kristensen2017adaptive,pronzato2012design}. For multi-fidelity simulations, where a high-cost high-accuracy and a low-cost low-accuracy simulators are available, a balance between the cost and accuracy can be achieved in designing the numerical simulations experiments~\cite{ghosh2019strategy}.

\begin{figure} [H]
\centering
 \subfigure[]{\includegraphics[width=\figwidthhalf,keepaspectratio]{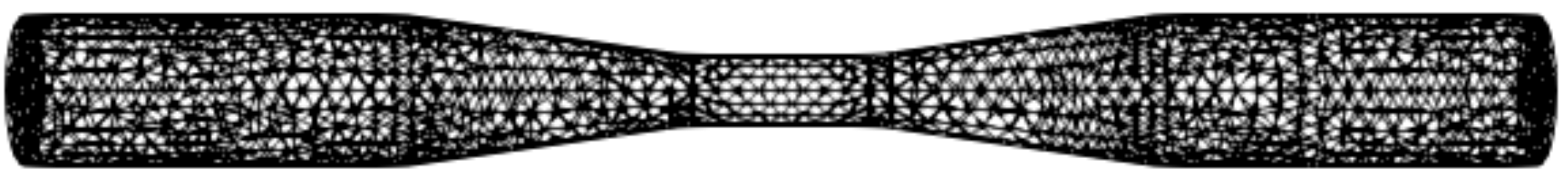}}
  \subfigure[]{\includegraphics[width=\figwidthhalf,keepaspectratio]{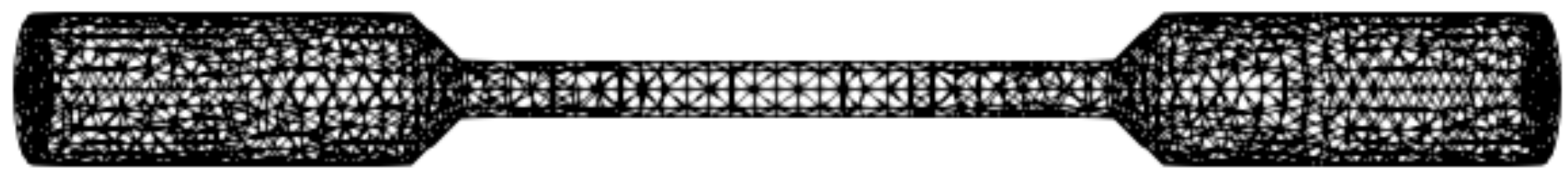}}
  \subfigure[]{\includegraphics[width=\figwidthhalf,keepaspectratio]{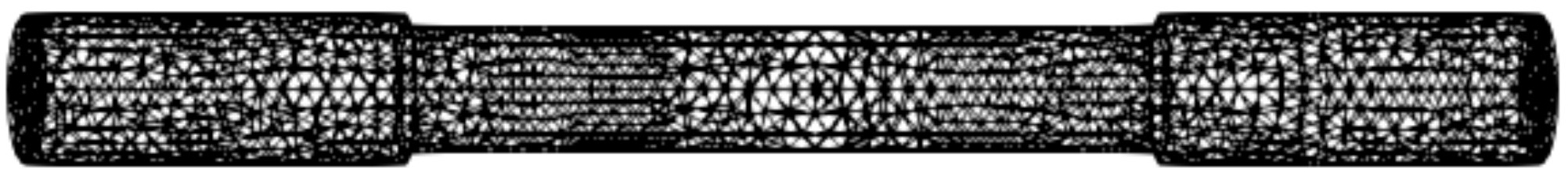}}
  \subfigure[]{\includegraphics[width=\figwidthhalf,keepaspectratio]{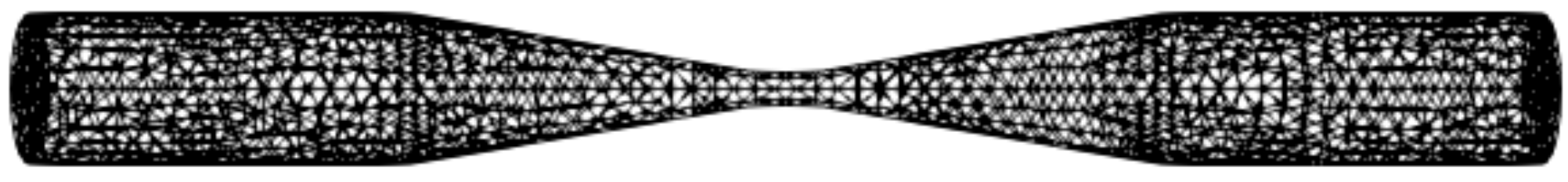}}
\caption{Four samples showing the variability in the beam geometry due to different values of the inner dimensions ($l_i,r_i$).}
 \label{samples}
 \end{figure}

\begin{figure} [H]
\centering
\includegraphics[width=\figwidthhalf,keepaspectratio]{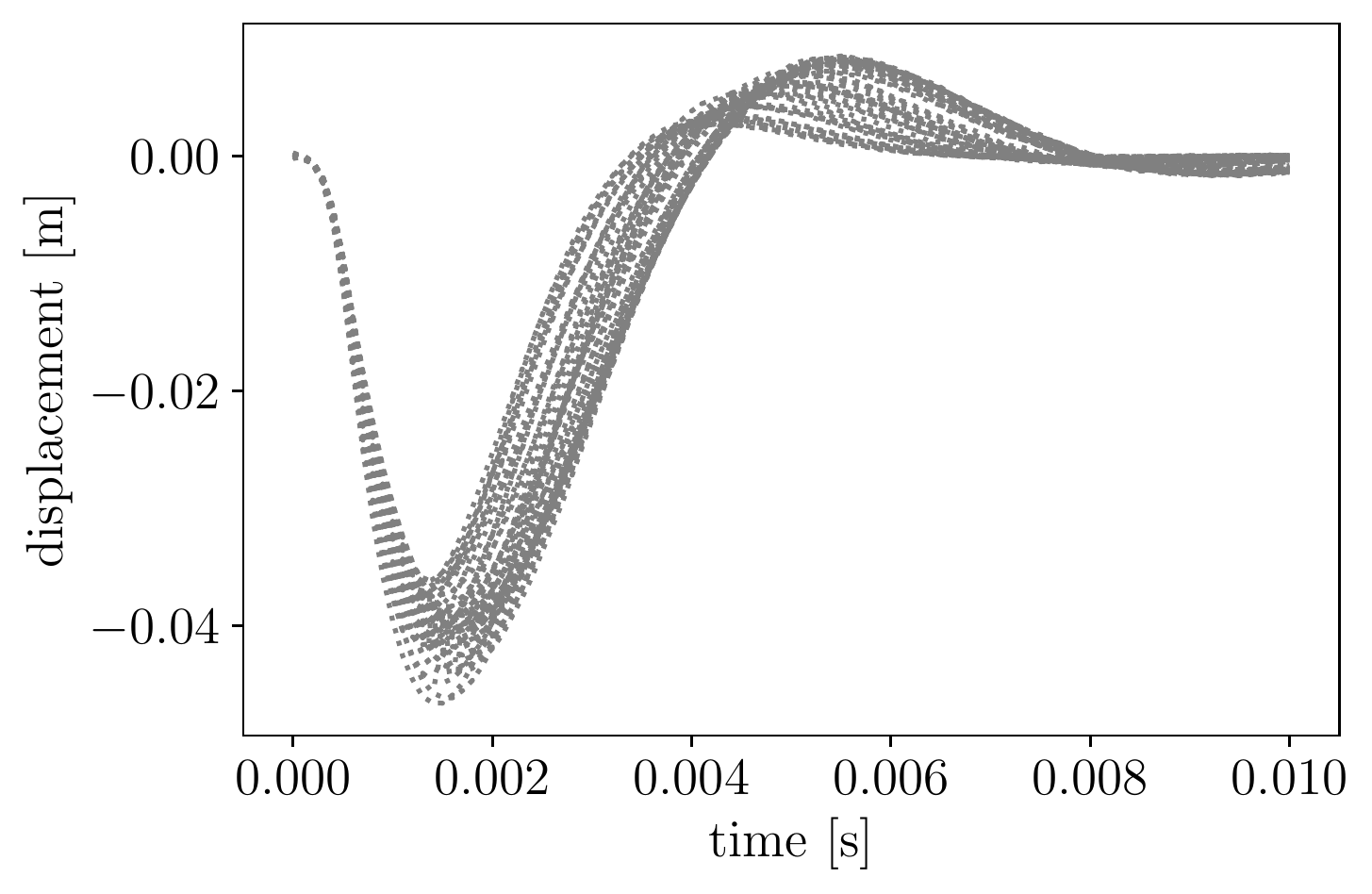}
\caption{The mid-span vertical displacement  of the 50 samples.}
\label{realizations}
 \end{figure}

The surrogate model is constructed based on samples that can  represent the variability in the beam geometry due to different values of the inner dimensions. We define the variability of the inner dimensions by assigning a uniform random distribution with a specified bounds as $l_i\sim U(0.25,0.75) {\rm m}$ and $r_i\sim U(0.01,0.05)  {\rm m}$. Using Latin hypercube sampling technique~\cite{lee2015pydoe}, we generate 50 independent samples for the inner dimensions. Using these samples, we generate the geometry of the beam followed by  constructing the corresponding finite element mesh, and executing the forward model to calculate the mid-span deflection (QoI). Samples of the training geometries are shown in Fig.~(\ref{samples}). Clearly, the samples span a wide range of the probable geometries of the beam. The corresponding scatter of the mid-span vertical displacement of the 50 samples are shown in Fig.~(\ref{realizations}). Of course, the variability of the inner dimensions not only affect the geometry, but also the location and magnitude of the bouncing deflection at around time $t=0.002 s$ and  $t=0.005 s$.

\begin{figure} [H]
\centering
 \subfigure[]{\includegraphics[width=\figwidthhalf,keepaspectratio]{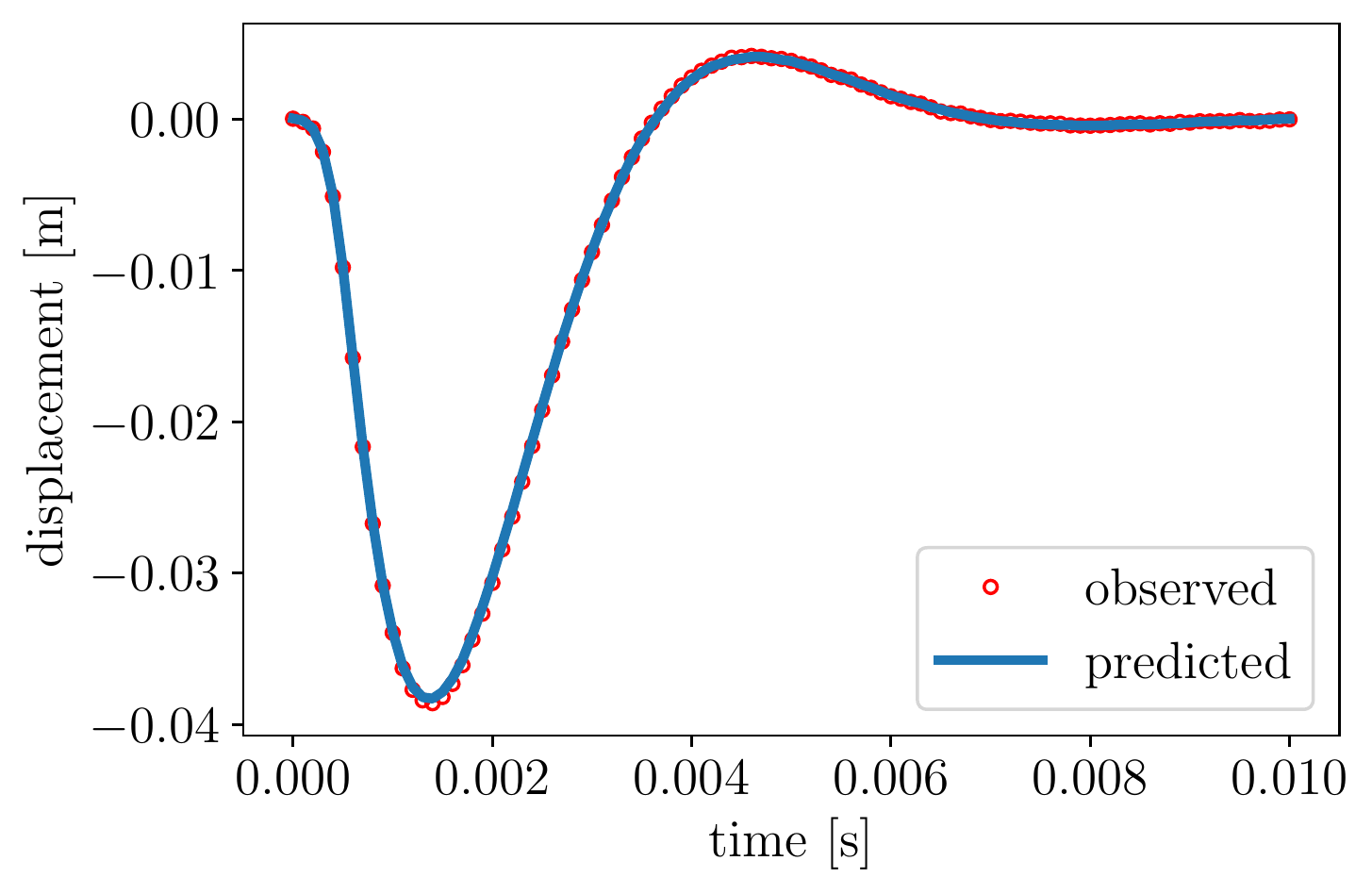}}
  \subfigure[]{\includegraphics[width=\figwidthhalf,keepaspectratio]{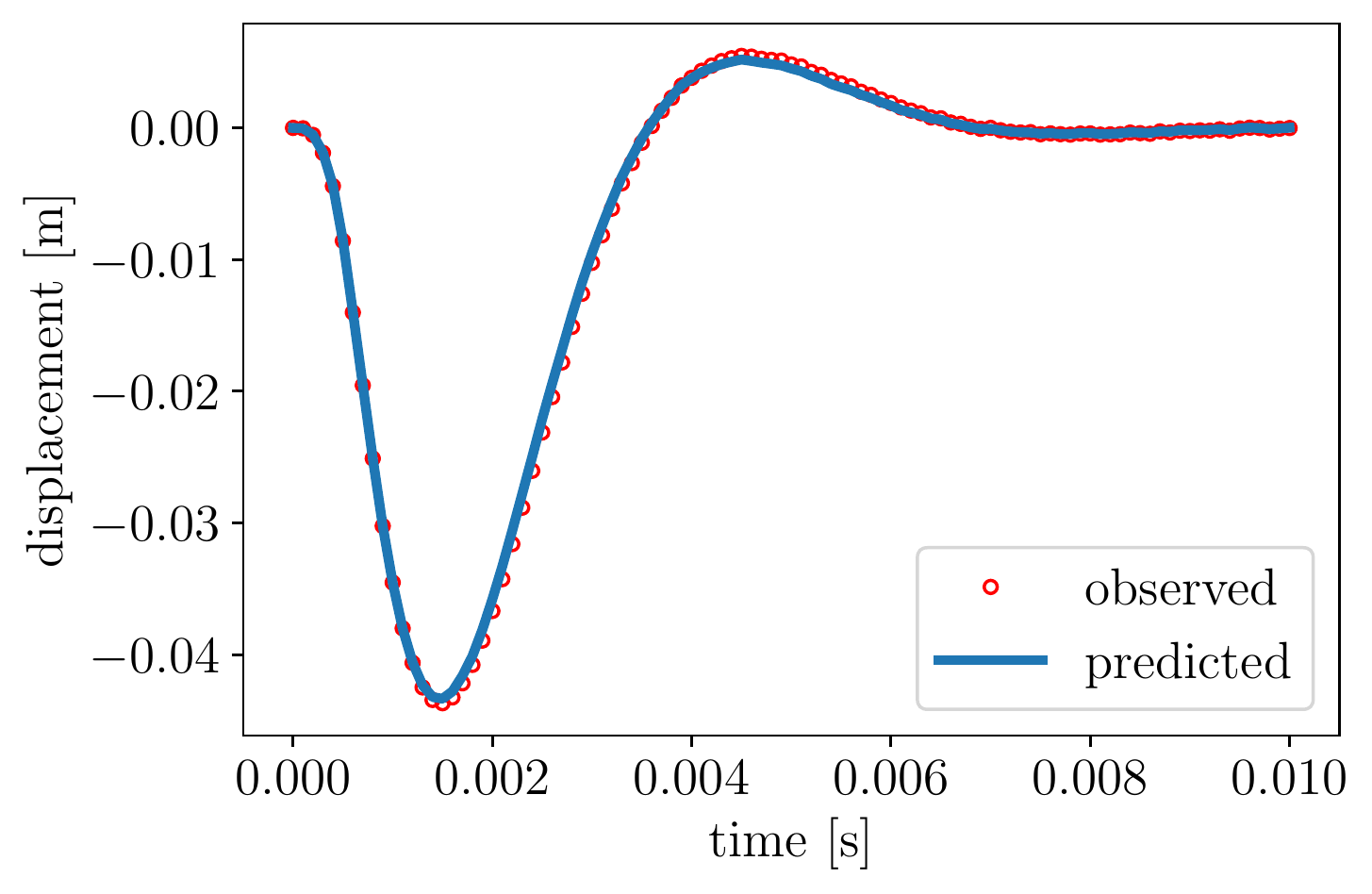}}
  \subfigure[]{\includegraphics[width=\figwidthhalf,keepaspectratio]{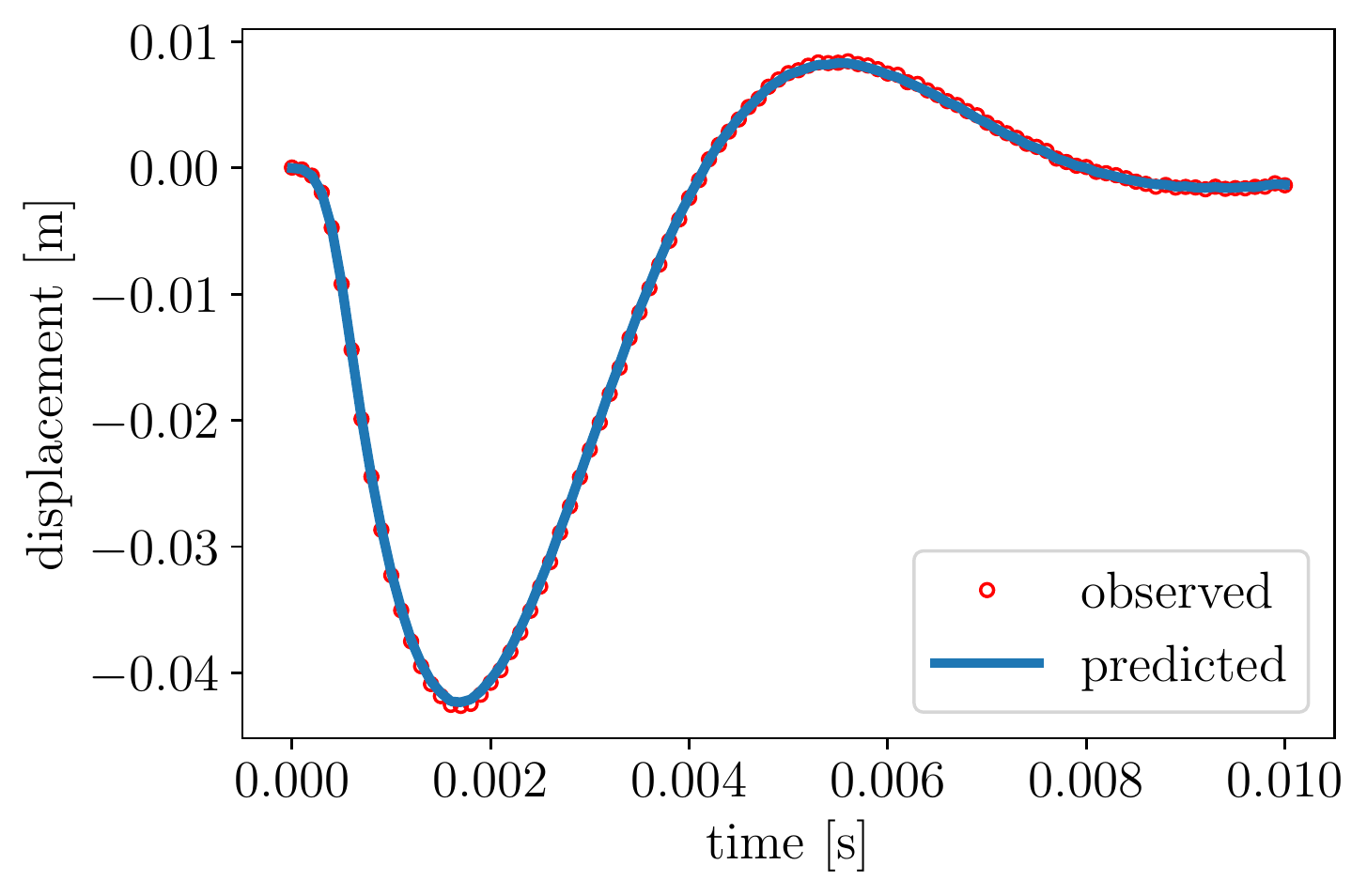}}
  \subfigure[]{\includegraphics[width=\figwidthhalf,keepaspectratio]{figs/test_one_gp_5.pdf}}
\caption{Observed and predicted QoI for different testing samples. The test samples are not part of the training set.}
\label{testing_samples}
\end{figure}

We randomly split the 50 samples into 40 samples for training and 10 for testing. For practical numerical implementation and to reflect the reality of the real world, we add a Gaussian random noise of strength $10^{-3} \times \max(u)$ to the deflection measurements $u$. The GP surrogate model is trained on the training samples and used to predict the held-out testing samples. Fig.(~\ref{testing_samples}) shows samples of observed and predicted responses for different values of the inner dimensions. The maximum and minutemen values of the mean squared error between the prediction and the observed response are $2.10 \times 10^{-7}$ and  $5.35 \times 10^{-9}$, respectively. Given the fact that the testing samples are not seen by the model during the training phase, the GP model can predict the unseen data within the given accuracy.

\begin{figure} [H]
\centering
\includegraphics[width=\figwidthhalf,keepaspectratio]{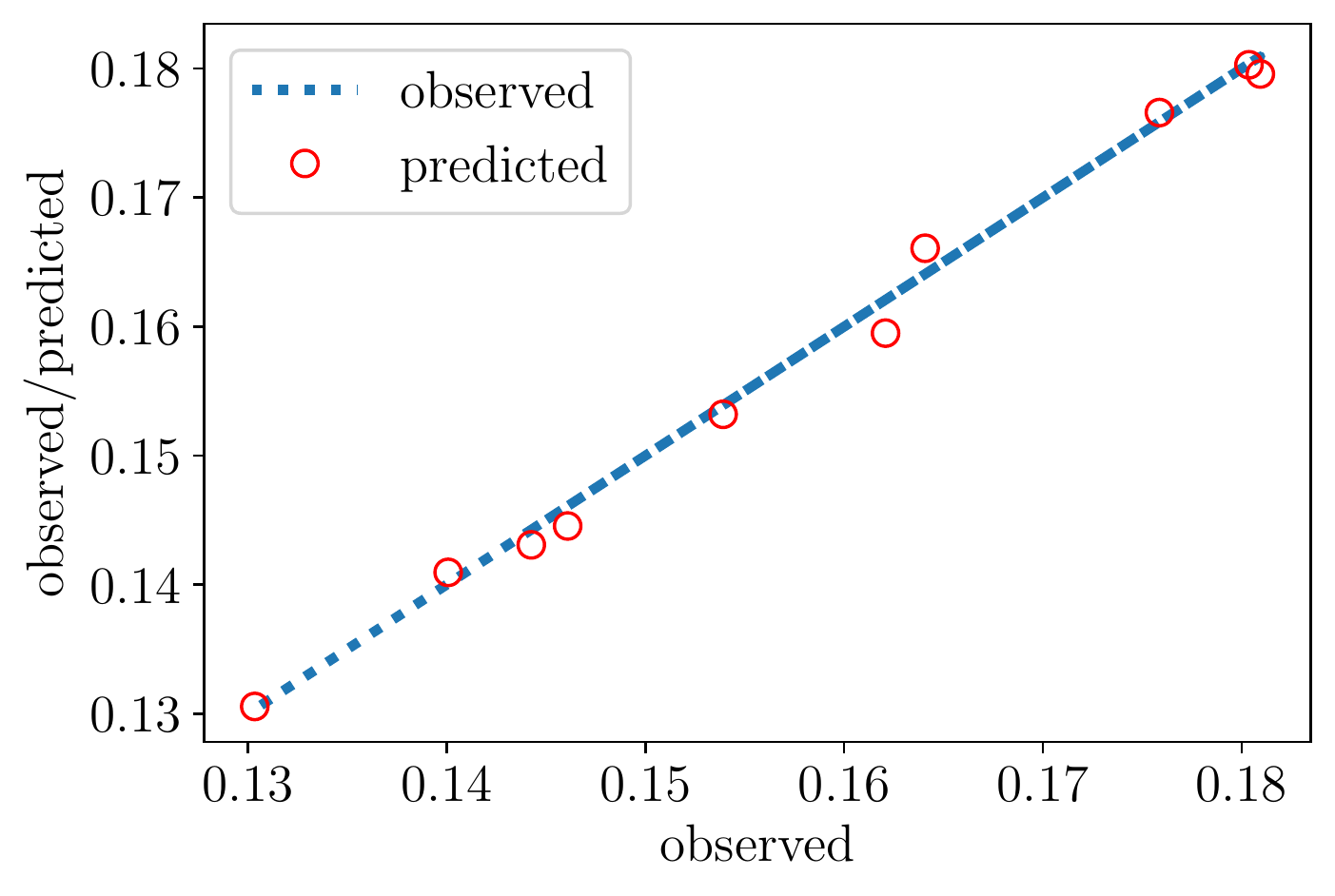}
\caption{The observed/predicted validation plot showing the norm of the observed (test data) and the corresponding model predictions.}
\label{norms}
 \end{figure}

To summarize the quality of the prediction, in Fig.~(\ref{norms}), we show the $L_2$-norm of the observed and predicted QoI. The observed/predicted validation plot indicates that the coefficient of determination between the  prediction and observation is $0.98$, and the corresponding  mean squared error is $2.53\times10 ^{-6}$. These statistical metrics indicate that the GP model can estimate the unseen geometry from a noisy measurement of the QoI within a given accuracy.

Once the GP model is validated, it can be deployed as a low-cost surrogate for the 3D finite element analysis code. The prediction of GP model takes only a fraction of the time that is needed by the finite element code to estimate the QoI with a fair accuracy.
 
\subsection{The Backward Problem}
\label{IP}
In the backward problem, we try to estimate the inner dimensions ($l_i,r_i$) of the beam from noisy measurements of the QoI. 
To this end, we utilize the GP surrogate model constructed in the previous subsection as a substitute for the forward model within the Bayesian framework. 

\begin{figure} [H]
\centering
\includegraphics[width=\figwidthhalf,keepaspectratio]{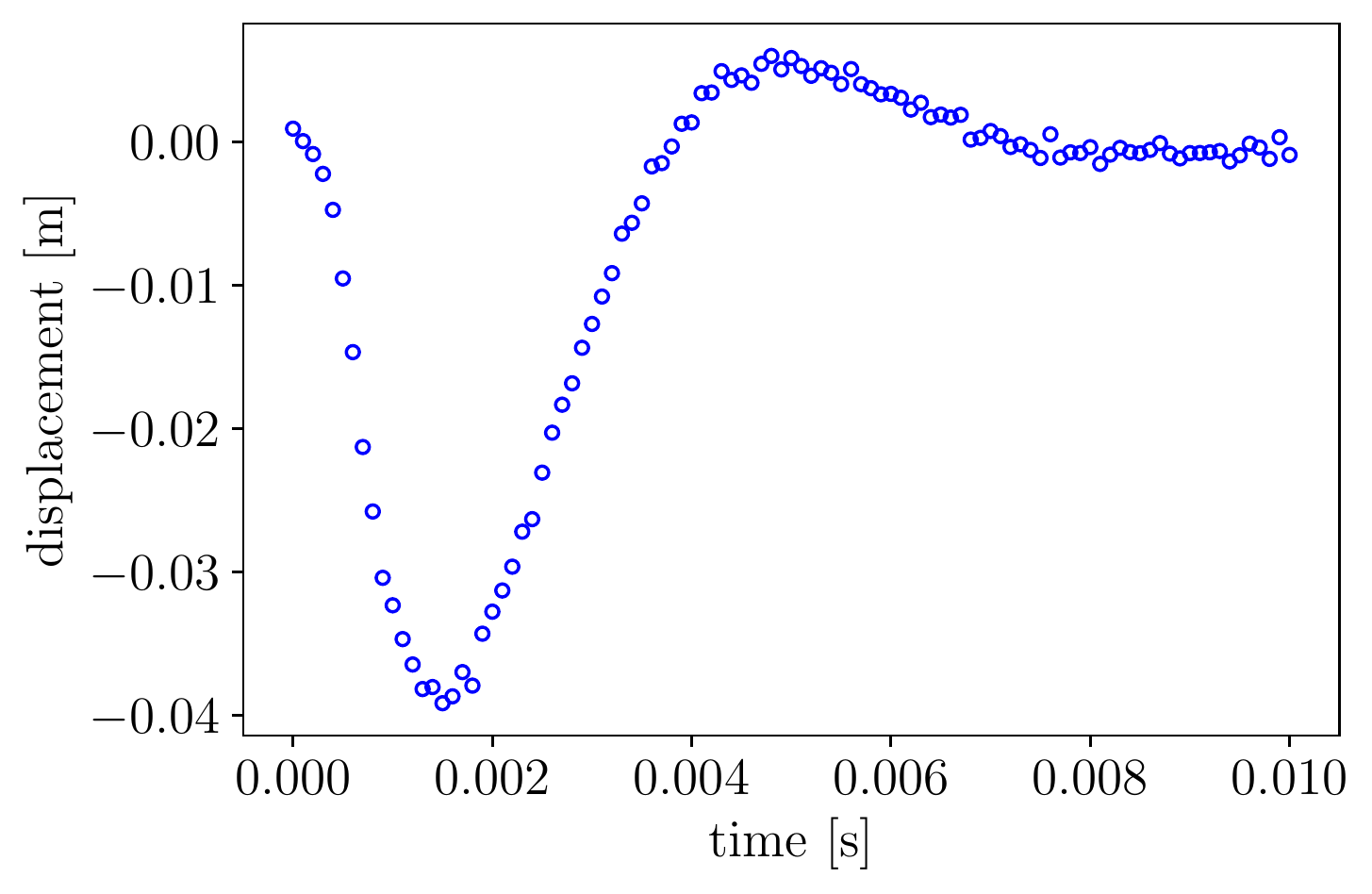}
\caption{Noisy measurement of the QoI }
\label{measurement}
   \end{figure}

We assume that a noisy measurement  for the QoI is available as shown in Fig.~(\ref{measurement}). The synthetic data is generated using inner dimension $l_i=0.313$ m and $r_i=0.055$ m plus ($\sigma_n=0.1 \times \max(u)$) Gaussian noise to mimic a real experiment setting.  

\begin{figure} [H]
\centering
  \subfigure[the inner length $l_i$ ]{\includegraphics[width=\figwidthhalf,keepaspectratio]{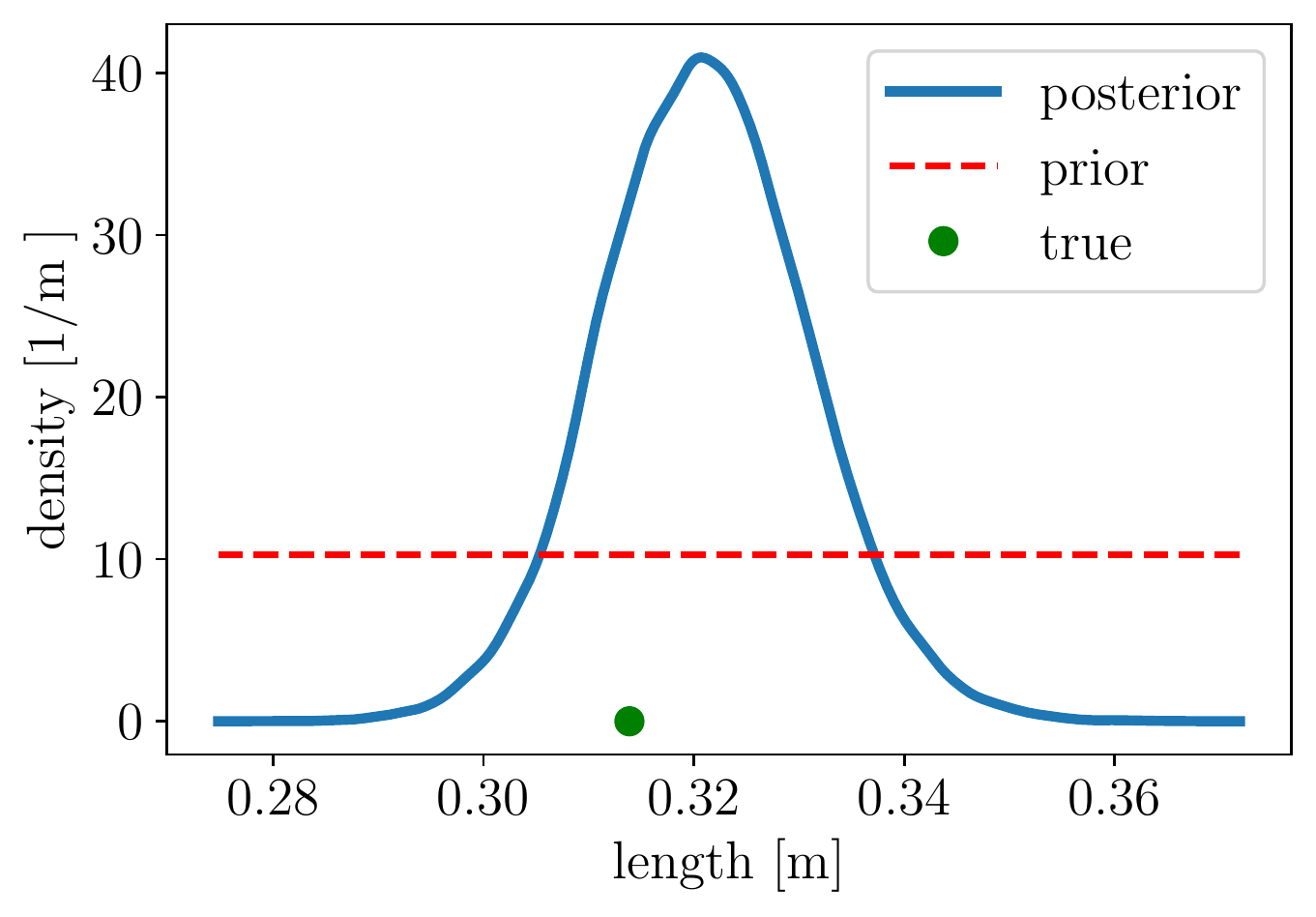}}
  \subfigure[the inner radius $r_i$]{\includegraphics[width=\figwidthhalf,keepaspectratio]{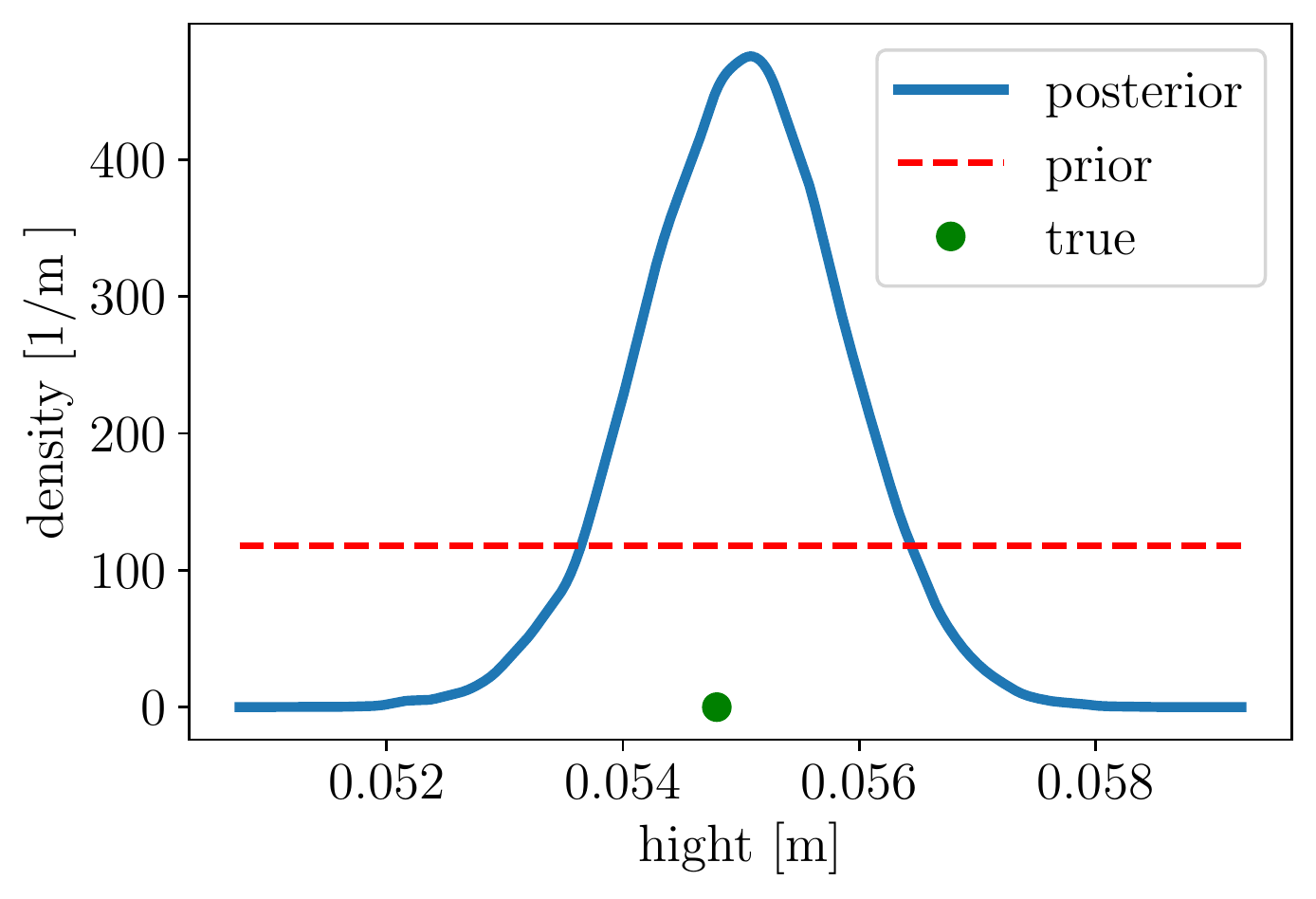}}
\caption{The estimated posterior density function of the inner dimensions  $\theta=[l_i, r_i]$. The sold line is the posterior PDF, the dotted line is the prior PDF and the bullet dot represents the true value  $l_i=0.313$ m and $r_i=0.055$ m.}
\label{posterior}
 \end{figure}

For the Bayesian calculation, we use non-informative prior for both the parameters $\theta=[l_i, r_i]$ and utilize an adaptive MCMC method (DRAM)~\cite{Haario2006,miles2019pymcmcstat} to estimate the posterior density.  In Fig.~(\ref{posterior}), we show the estimated posterior density of  the parameters $\theta=[l_i, r_i]$ . We also show the prior density and the true value of the parameters. Note that the true parameters where not part of  either the training nor the testing datasets. This highlights the robustness of the framework. The  mean of the estimated values  are $l_i=0.310\pm 0.048$ m and $r_i=0.054\pm 0.004$ m (the confidence bounds are based on two standard deviation). 

\begin{figure} [H]
\centering
\includegraphics[width=\figwidthhalf,keepaspectratio]{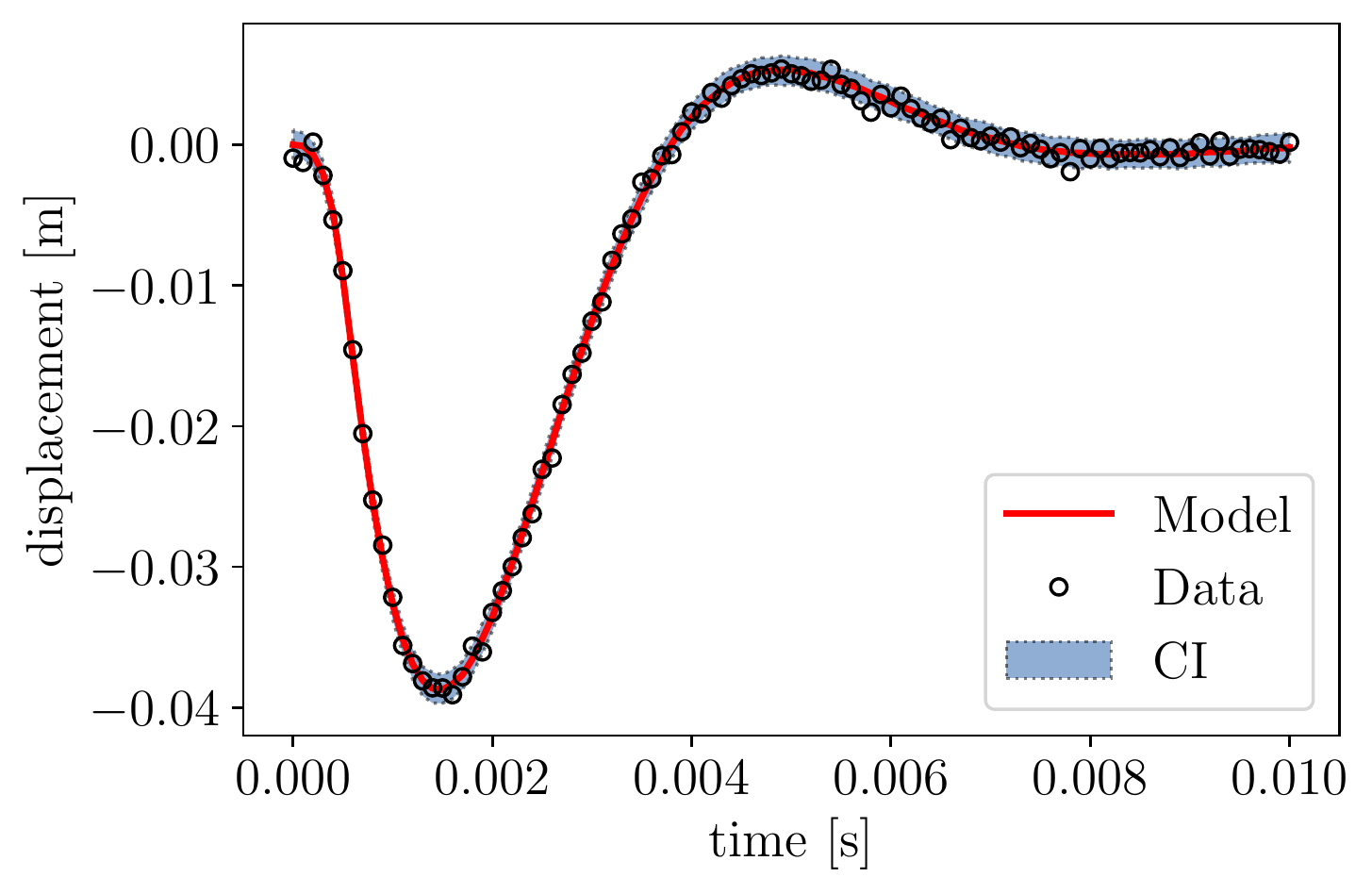}
  \caption{The prediction of the surrogate model and its confidence interval due to uncertainty propagation of the variability in the estimated inner dimensions.}
  \label{push}
   \end{figure}

Next, the uncertainty in the parameter estimation represented by the posterior density in Fig.~(\ref{posterior}) is propagated forward through the surrogate model to estimate a confidence bound on the prediction of the QoI. In Fig.~(\ref{push}), we show the model prediction and the $95\%$ confidence interval  as well as the true measured response. The $L_2$ for the discrepancy between the mean model prediction and the measured data is $0.005$ m. This conforms that the response due to the estimated parameters uncertainty agrees reasonably well with the true response.
 
\subsection{Localized Uncertainty Propagation} 
\begin{figure} [H]
\centering 
  \subfigure[displacement]{\includegraphics[width=\figwidthhalf,keepaspectratio]{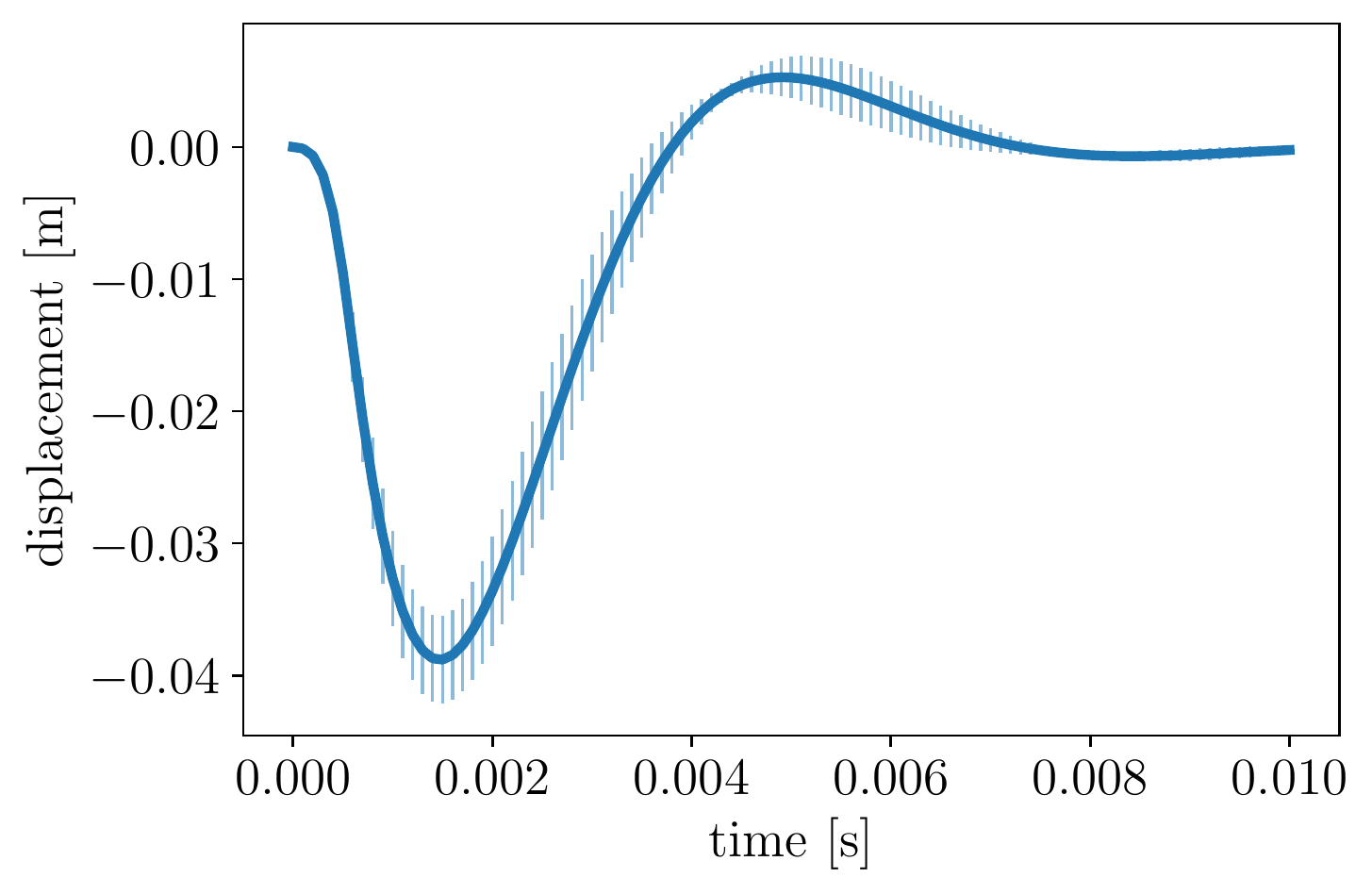}}
  \subfigure[velocity]{\includegraphics[width=\figwidthhalf,keepaspectratio]{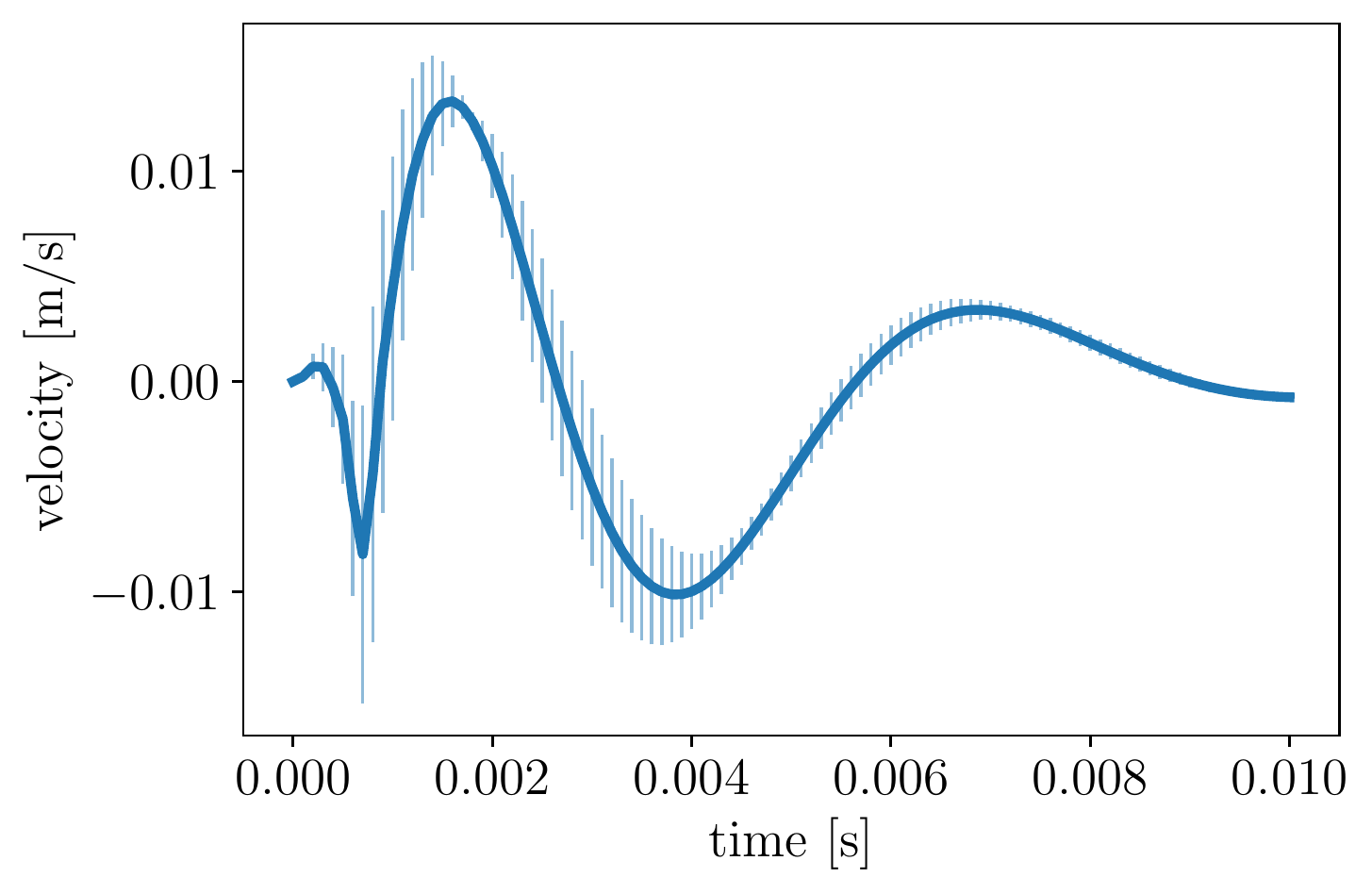}}
  \caption{The PC prediction of the  displacement and velocity at the mid-span. The uncertainty bounds represent two standard deviation.}
\label{PC_local}
 \end{figure}

 The QoI is confined within the core cylinder defined by inner dimensions $\theta=[l_i, r_i]$.  Once these dimensions are available, the effect of the random variability in the material properties of the  inner subdomain can be estimated using PC expansion.  Without loss of generality, here we assume that  for the inner cylinder, the Young's modulus and material density are random quantities, while  Poisson's ratio  is deterministic as
\begin{align}
 E({\bf x}, \xi_1)= 
\begin{cases}
    E_0 (1+\sigma_E \xi_1) ,& \text{for } {\bf x} \in \Omega_2\\
    E_0,              & \text{otherwise}
\end{cases}
\end{align} 
 \noindent
 and 
 \begin{align}
\rho({\bf x}, \xi_2)= 
\begin{cases}
   \rho_0(1+\sigma_\rho~\xi_2),& \text{for } {\bf x} \in \Omega_2\\
    \rho_0,              & \text{otherwise}
\end{cases}
\end{align} 
 \noindent
where the artificial boundary for $ \Omega_2$ are defined by the Maximum A Posteriori (MAP) estimation  of the  inner dimensions $\theta=[l_i, r_i]$, $E_0=70$ GPa, $\rho_0=26.25$ kN/$\rm m^3$, $\sigma_E =0.25$ and $\sigma_\rho=0.15$ and $\xi_1, \xi_2$ are standard normal random variables. Note that, not only the solution over $\Omega_2$ is stochastic, but also over all the whole domain since the spatial finite element and stochastic basis functions are continuous across the domains interfaces. We use second order PC expansion to propagate the localized uncertainty due to the random Young's modulus and material density as shown in Fig.~(\ref{PC_local}). The uncertainty bounds follow the trend of the response, with a higher value near the shock location.  Although not explored here, high  spatio-temporal resolution  solver can be directed toward the region of interest, while a less resolution alternative can be assigned to  the regions away from the QoI. As demonstrated in~\cite{subber2017asynchronous,subber2018uncertainty,subber2016asynchronous}, PASTA-DDM-UQ approach leads to a customized solver for localized uncertainty propagation with less computational cost.

 \section{Conclusion}
\label{Conclusion}
We present a data-based partitioning scheme  for localized uncertainty quantification in elastodynamic system. The localized region of interest is identified using  Bayesian inference framework. Measurement of the system response at one location in conjunction with a physics-based computational model is used to infer the localized features of the region of interested. A data-based surrogate model for the physics-based simulator is constructed using Gaussian process regression in order to reduce the computational cost of the Bayesian inversion. Material uncertainty in the region of interest is propagated through the system using polynomial chaos. We exercise our framework on a three-dimensional beam with localized feature and subjected to an impact load. The presented framework can facilitate quantifying the effect of the sub-component uncertainty on the system-level. Proper assessment of uncertainty  at various level can accelerate the adaptation process of a new component introduced to an existing system.

\bibliographystyle{unsrt}
\bibliography{ref}

\end{document}